
%
%
%
%
%
\def\standardrisposta{s }\def\reducedrisposta{r }
\def\mplarisposta{mpla }\def\zerorisposta{z }
\def\doublerisposta{d }\def\cartarisposta{e }\def\amsrisposta{y }
\newcount\ingrandimento \newcount\sinnota \newcount\dimnota
\newcount\unoduecol \newdimen\collhsize \newdimen\tothsize
\newdimen\fullhsize \newcount\controllorisposta \sinnota=1
\newskip\infralinea  \global\controllorisposta=0
\immediate\write16 { ********  Welcome to PANDA macros (Plain TeX,
AP, 1991) ******** }
\immediate\write16 { You'll have to answer a few questions in
lowercase.}
\message{>  Do you want it in double-page (d), reduced (r)
or standard format (s) ? }
\message{>  Do you want it in USA A4 (u) or EUROPEAN A4
(e) paper size ? }
\message{>  Do you have AMSFonts 2.0 (math) fonts (y/n) ? }
%
%
%
%
%
\ifx\risposta\standardrisposta \ingrandimento=1200
\message {>> This will come out UNREDUCED << }
\dimnota=2 \unoduecol=1 \global\controllorisposta=1 \fi
\ifx\risposta\reducedrisposta \ingrandimento=1095 \dimnota=1
\unoduecol=1  \global\controllorisposta=1
\message {>> This will come out REDUCED << } \fi
\ifx\risposta\doublerisposta \ingrandimento=1000 \dimnota=2
\unoduecol=2  \message {>> You must print this in
LANDSCAPE orientation << } \global\controllorisposta=1 \fi
\ifx\risposta\mplarisposta \ingrandimento=1000 \dimnota=1
\message {>> Mod. Phys. Lett. A format << }
\unoduecol=1 \global\controllorisposta=1 \fi
\ifx\risposta\zerorisposta \ingrandimento=1000 \dimnota=2
\message {>> Zero Magnification format << }
\unoduecol=1 \global\controllorisposta=1 \fi
\ifnum\controllorisposta=0  \ingrandimento=1200
\message {>>> ERROR IN INPUT, I ASSUME STANDARD
UNREDUCED FORMAT <<< }  \dimnota=2 \unoduecol=1 \fi
\magnification=\ingrandimento
%
%
%
%
\newdimen\eucolumnsize \newdimen\eudoublehsize \newdimen\eudoublevsize
\newdimen\uscolumnsize \newdimen\usdoublehsize \newdimen\usdoublevsize
\newdimen\eusinglehsize \newdimen\eusinglevsize \newdimen\ussinglehsize
\newskip\standardbaselineskip \newdimen\ussinglevsize
\newskip\reducedbaselineskip \newskip\doublebaselineskip
\eucolumnsize=12.0truecm    
\eudoublehsize=25.5truecm   
\eudoublevsize=6.5truein    
\uscolumnsize=4.4truein     
\usdoublehsize=9.4truein    
\usdoublevsize=6.8truein    
\eusinglehsize=6.5truein    
\eusinglevsize=24truecm     
\ussinglehsize=6.5truein    
\ussinglevsize=8.9truein    
\standardbaselineskip=16pt plus.2pt  
\reducedbaselineskip=14pt plus.2pt   
\doublebaselineskip=12pt plus.2pt    
%
%
\def\Portoffset{}
\def\Landoffset{\hoffset=-.140truein\voffset=-.20truein}
\ifx\risposta\mplarisposta \def\Portoffset{\hoffset=1.9truecm
\voffset=1.4truecm} \fi
%
%
\def\Landspec{ }
\tolerance=10000
\parskip=0pt plus2pt  \leftskip=0pt \rightskip=0pt
%
%
\ifx\risposta\standardrisposta \infralinea=\standardbaselineskip \fi
\ifx\risposta\reducedrisposta  \infralinea=\reducedbaselineskip \fi
\ifx\risposta\doublerisposta   \infralinea=\doublebaselineskip \fi
\ifx\risposta\mplarisposta     \infralinea=13pt \fi
\ifx\risposta\zerorisposta     \infralinea=12pt plus.2pt\fi
\ifnum\controllorisposta=0    \infralinea=\standardbaselineskip \fi
\ifx\risposta\doublerisposta   \Landoffset \else \Portoffset \fi
\ifx\risposta\doublerisposta \ifx\srisposta\cartarisposta
\tothsize=\eudoublehsize \collhsize=\eucolumnsize
\vsize=\eudoublevsize  \else  \tothsize=\usdoublehsize
\collhsize=\uscolumnsize \vsize=\usdoublevsize \fi \else
\ifx\srisposta\cartarisposta \tothsize=\eusinglehsize
\vsize=\eusinglevsize \else  \tothsize=\ussinglehsize
\vsize=\ussinglevsize \fi \collhsize=4.4truein \fi
\ifx\risposta\mplarisposta \tothsize=5.0truein
\vsize=7.8truein \collhsize=4.4truein \fi
%
%
%
%
\newcount\contaeuler \newcount\contacyrill \newcount\contaams
\font\ninerm=cmr9  \font\eightrm=cmr8  \font\sixrm=cmr6
\font\ninei=cmmi9  \font\eighti=cmmi8  \font\sixi=cmmi6
\font\ninesy=cmsy9  \font\eightsy=cmsy8  \font\sixsy=cmsy6
\font\ninebf=cmbx9  \font\eightbf=cmbx8  \font\sixbf=cmbx6
\font\ninett=cmtt9  \font\eighttt=cmtt8  \font\nineit=cmti9
\font\eightit=cmti8 \font\ninesl=cmsl9  \font\eightsl=cmsl8
\skewchar\ninei='177 \skewchar\eighti='177 \skewchar\sixi='177
\skewchar\ninesy='60 \skewchar\eightsy='60 \skewchar\sixsy='60
\hyphenchar\ninett=-1 \hyphenchar\eighttt=-1 \hyphenchar\tentt=-1
%
\font\tencmmib=cmmib10  \newfam\cmmibfam  \skewchar\tencmmib='177
\font\tencmbsy=cmbsy10  \newfam\cmbsyfam  \skewchar\tencmbsy='60
\def\scaps{\cmcsc}                 
\font\tencmcsc=cmcsc10  \newfam\cmcscfam
\ifnum\ingrandimento=1095

\font\capsone=cmcsc10 at 10.95pt 

\else

\font\capsone=cmcsc10 at 12pt 
\fi
\def\chapterfont#1{\xdef\ttaarr{#1}}
\def\sectionfont#1{\xdef\ppaarr{#1}}
\def\ttaarr{\bf}		
\def\ppaarr{\sl}		

%
%
%
\newfam\eufmfam \newfam\msamfam \newfam\msbmfam \newfam\eufbfam
\def\Loadeulerfonts{\global\contaeuler=1 \ifx\arisposta\amsrisposta
\font\teneufm=eufm10              
\font\eighteufm=eufm8 \font\nineeufm=eufm9 \font\sixeufm=eufm6
\font\seveneufm=eufm7  \font\fiveeufm=eufm5
\font\teneufb=eufb10              
\font\eighteufb=eufb8 \font\nineeufb=eufb9 \font\sixeufb=eufb6
\font\seveneufb=eufb7  \font\fiveeufb=eufb5
\font\teneurm=eurm10              
\font\eighteurm=eurm8 \font\nineeurm=eurm9
\font\teneurb=eurb10              
\font\eighteurb=eurb8 \font\nineeurb=eurb9
\font\teneusm=eusm10              
\font\eighteusm=eusm8 \font\nineeusm=eusm9
\font\teneusb=eusb10              
\font\eighteusb=eusb8 \font\nineeusb=eusb9
\else \def\eufm{\tt} \def\eufb{\tt} \def\eurm{\tt} \def\eurb{\tt}
\def\eusm{\tt} \def\eusb{\tt}    \fi}

\def\loadamsmath{\global\contaams=1 \ifx\arisposta\amsrisposta
\font\tenmsam=msam10 \font\ninemsam=msam9 \font\eightmsam=msam8
\font\sevenmsam=msam7 \font\sixmsam=msam6 \font\fivemsam=msam5
\font\tenmsbm=msbm10 \font\ninemsbm=msbm9 \font\eightmsbm=msbm8
\font\sevenmsbm=msbm7 \font\sixmsbm=msbm6 \font\fivemsbm=msbm5
\else \def\msbm{\bf} \fi \def\Bbb{\msbm} \def\symbl{\msam} \tenpoint}
\def\loadcyrill{\global\contacyrill=1 \ifx\arisposta\amsrisposta
\font\tenwncyr=wncyr10 \font\ninewncyr=wncyr9 \font\eightwncyr=wncyr8
\font\tenwncyb=wncyr10 \font\ninewncyb=wncyr9 \font\eightwncyb=wncyr8
\font\tenwncyi=wncyr10 \font\ninewncyi=wncyr9 \font\eightwncyi=wncyr8
\else \def\cyrill{\sl} \def\cyrilb{\sl} \def\cyrili{\sl} \fi\tenpoint}
\ifx\arisposta\amsrisposta
\font\sevenex=cmex7               
\font\eightex=cmex8  \font\nineex=cmex9
\font\ninecmmib=cmmib9   \font\eightcmmib=cmmib8
\font\sevencmmib=cmmib7 \font\sixcmmib=cmmib6
\font\fivecmmib=cmmib5   \skewchar\ninecmmib='177
\skewchar\eightcmmib='177  \skewchar\sevencmmib='177
\skewchar\sixcmmib='177   \skewchar\fivecmmib='177
\font\ninecmbsy=cmbsy9    \font\eightcmbsy=cmbsy8
\font\sevencmbsy=cmbsy7  \font\sixcmbsy=cmbsy6
\font\fivecmbsy=cmbsy5   \skewchar\ninecmbsy='60
\skewchar\eightcmbsy='60  \skewchar\sevencmbsy='60
\skewchar\sixcmbsy='60    \skewchar\fivecmbsy='60
\font\ninecmcsc=cmcsc9    \font\eightcmcsc=cmcsc8     \else
\def\cmmib{\fam\cmmibfam\tencmmib}\textfont\cmmibfam=\tencmmib
\scriptfont\cmmibfam=\tencmmib \scriptscriptfont\cmmibfam=\tencmmib
\def\cmbsy{\fam\cmbsyfam\tencmbsy} \textfont\cmbsyfam=\tencmbsy
\scriptfont\cmbsyfam=\tencmbsy \scriptscriptfont\cmbsyfam=\tencmbsy
\scriptfont\cmcscfam=\tencmcsc \scriptscriptfont\cmcscfam=\tencmcsc
\def\cmcsc{\fam\cmcscfam\tencmcsc} \textfont\cmcscfam=\tencmcsc \fi
\catcode`@=11
\newskip\ttglue
\gdef\tenpoint{\def\rm{\fam0\tenrm}
  \textfont0=\tenrm \scriptfont0=\sevenrm \scriptscriptfont0=\fiverm
  \textfont1=\teni \scriptfont1=\seveni \scriptscriptfont1=\fivei
  \textfont2=\tensy \scriptfont2=\sevensy \scriptscriptfont2=\fivesy
  \textfont3=\tenex \scriptfont3=\tenex \scriptscriptfont3=\tenex
  \def\mcal{\fam2 \tensy}  \def\mmit{\fam1 \teni}
  \textfont\itfam=\tenit \def\it{\fam\itfam\tenit}
  \textfont\slfam=\tensl \def\sl{\fam\slfam\tensl}
  \textfont\ttfam=\tentt \scriptfont\ttfam=\eighttt
  \scriptscriptfont\ttfam=\eighttt  \def\tt{\fam\ttfam\tentt}
  \textfont\bffam=\tenbf \scriptfont\bffam=\sevenbf
  \scriptscriptfont\bffam=\fivebf \def\bf{\fam\bffam\tenbf}
     \ifx\arisposta\amsrisposta    \ifnum\contaeuler=1
  \textfont\eufmfam=\teneufm \scriptfont\eufmfam=\seveneufm
  \scriptscriptfont\eufmfam=\fiveeufm \def\eufm{\fam\eufmfam\teneufm}
  \textfont\eufbfam=\teneufb \scriptfont\eufbfam=\seveneufb
  \scriptscriptfont\eufbfam=\fiveeufb \def\eufb{\fam\eufbfam\teneufb}
  \def\eurm{\teneurm} \def\eurb{\teneurb} \def\eusm{\teneusm}
  \def\eusb{\teneusb}    \fi    \ifnum\contaams=1
  \textfont\msamfam=\tenmsam \scriptfont\msamfam=\sevenmsam
  \scriptscriptfont\msamfam=\fivemsam \def\msam{\fam\msamfam\tenmsam}
  \textfont\msbmfam=\tenmsbm \scriptfont\msbmfam=\sevenmsbm
  \scriptscriptfont\msbmfam=\fivemsbm \def\msbm{\fam\msbmfam\tenmsbm}
     \fi      \ifnum\contacyrill=1     \def\cyrill{\tenwncyr}
  \def\cyrilb{\tenwncyb}  \def\cyrili{\tenwncyi}         \fi
  \textfont3=\tenex \scriptfont3=\sevenex \scriptscriptfont3=\sevenex
  \def\cmmib{\fam\cmmibfam\tencmmib} \scriptfont\cmmibfam=\sevencmmib
  \textfont\cmmibfam=\tencmmib  \scriptscriptfont\cmmibfam=\fivecmmib
  \def\cmbsy{\fam\cmbsyfam\tencmbsy} \scriptfont\cmbsyfam=\sevencmbsy
  \textfont\cmbsyfam=\tencmbsy  \scriptscriptfont\cmbsyfam=\fivecmbsy
  \def\cmcsc{\fam\cmcscfam\tencmcsc} \scriptfont\cmcscfam=\eightcmcsc
  \textfont\cmcscfam=\tencmcsc \scriptscriptfont\cmcscfam=\eightcmcsc
     \fi            \tt \ttglue=.5em plus.25em minus.15em
  \normalbaselineskip=12pt
  \setbox\strutbox=\hbox{\vrule height8.5pt depth3.5pt width0pt}
  \let\sc=\eightrm \let\big=\tenbig   \normalbaselines
  \baselineskip=\infralinea  \rm}
\gdef\ninepoint{\def\rm{\fam0\ninerm}
  \textfont0=\ninerm \scriptfont0=\sixrm \scriptscriptfont0=\fiverm
  \textfont1=\ninei \scriptfont1=\sixi \scriptscriptfont1=\fivei
  \textfont2=\ninesy \scriptfont2=\sixsy \scriptscriptfont2=\fivesy
  \textfont3=\tenex \scriptfont3=\tenex \scriptscriptfont3=\tenex
  \def\mcal{\fam2 \ninesy}  \def\mmit{\fam1 \ninei}
  \textfont\itfam=\nineit \def\it{\fam\itfam\nineit}
  \textfont\slfam=\ninesl \def\sl{\fam\slfam\ninesl}
  \textfont\ttfam=\ninett \scriptfont\ttfam=\eighttt
  \scriptscriptfont\ttfam=\eighttt \def\tt{\fam\ttfam\ninett}
  \textfont\bffam=\ninebf \scriptfont\bffam=\sixbf
  \scriptscriptfont\bffam=\fivebf \def\bf{\fam\bffam\ninebf}
     \ifx\arisposta\amsrisposta  \ifnum\contaeuler=1
  \textfont\eufmfam=\nineeufm \scriptfont\eufmfam=\sixeufm
  \scriptscriptfont\eufmfam=\fiveeufm \def\eufm{\fam\eufmfam\nineeufm}
  \textfont\eufbfam=\nineeufb \scriptfont\eufbfam=\sixeufb
  \scriptscriptfont\eufbfam=\fiveeufb \def\eufb{\fam\eufbfam\nineeufb}
  \def\eurm{\nineeurm} \def\eurb{\nineeurb} \def\eusm{\nineeusm}
  \def\eusb{\nineeusb}     \fi   \ifnum\contaams=1
  \textfont\msamfam=\ninemsam \scriptfont\msamfam=\sixmsam
  \scriptscriptfont\msamfam=\fivemsam \def\msam{\fam\msamfam\ninemsam}
  \textfont\msbmfam=\ninemsbm \scriptfont\msbmfam=\sixmsbm
  \scriptscriptfont\msbmfam=\fivemsbm \def\msbm{\fam\msbmfam\ninemsbm}
     \fi       \ifnum\contacyrill=1     \def\cyrill{\ninewncyr}
  \def\cyrilb{\ninewncyb}  \def\cyrili{\ninewncyi}         \fi
  \textfont3=\nineex \scriptfont3=\sevenex \scriptscriptfont3=\sevenex
  \def\cmmib{\fam\cmmibfam\ninecmmib}  \textfont\cmmibfam=\ninecmmib
  \scriptfont\cmmibfam=\sixcmmib \scriptscriptfont\cmmibfam=\fivecmmib
  \def\cmbsy{\fam\cmbsyfam\ninecmbsy}  \textfont\cmbsyfam=\ninecmbsy
  \scriptfont\cmbsyfam=\sixcmbsy \scriptscriptfont\cmbsyfam=\fivecmbsy
  \def\cmcsc{\fam\cmcscfam\ninecmcsc} \scriptfont\cmcscfam=\eightcmcsc
  \textfont\cmcscfam=\ninecmcsc \scriptscriptfont\cmcscfam=\eightcmcsc
     \fi            \tt \ttglue=.5em plus.25em minus.15em
  \normalbaselineskip=11pt
  \setbox\strutbox=\hbox{\vrule height8pt depth3pt width0pt}
  \let\sc=\sevenrm \let\big=\ninebig \normalbaselines\rm}
\gdef\eightpoint{\def\rm{\fam0\eightrm}
  \textfont0=\eightrm \scriptfont0=\sixrm \scriptscriptfont0=\fiverm
  \textfont1=\eighti \scriptfont1=\sixi \scriptscriptfont1=\fivei
  \textfont2=\eightsy \scriptfont2=\sixsy \scriptscriptfont2=\fivesy
  \textfont3=\tenex \scriptfont3=\tenex \scriptscriptfont3=\tenex
  \def\mcal{\fam2 \eightsy}  \def\mmit{\fam1 \eighti}
  \textfont\itfam=\eightit \def\it{\fam\itfam\eightit}
  \textfont\slfam=\eightsl \def\sl{\fam\slfam\eightsl}
  \textfont\ttfam=\eighttt \scriptfont\ttfam=\eighttt
  \scriptscriptfont\ttfam=\eighttt \def\tt{\fam\ttfam\eighttt}
  \textfont\bffam=\eightbf \scriptfont\bffam=\sixbf
  \scriptscriptfont\bffam=\fivebf \def\bf{\fam\bffam\eightbf}
     \ifx\arisposta\amsrisposta   \ifnum\contaeuler=1
  \textfont\eufmfam=\eighteufm \scriptfont\eufmfam=\sixeufm
  \scriptscriptfont\eufmfam=\fiveeufm \def\eufm{\fam\eufmfam\eighteufm}
  \textfont\eufbfam=\eighteufb \scriptfont\eufbfam=\sixeufb
  \scriptscriptfont\eufbfam=\fiveeufb \def\eufb{\fam\eufbfam\eighteufb}
  \def\eurm{\eighteurm} \def\eurb{\eighteurb} \def\eusm{\eighteusm}
  \def\eusb{\eighteusb}       \fi    \ifnum\contaams=1
  \textfont\msamfam=\eightmsam \scriptfont\msamfam=\sixmsam
  \scriptscriptfont\msamfam=\fivemsam \def\msam{\fam\msamfam\eightmsam}
  \textfont\msbmfam=\eightmsbm \scriptfont\msbmfam=\sixmsbm
  \scriptscriptfont\msbmfam=\fivemsbm \def\msbm{\fam\msbmfam\eightmsbm}
     \fi       \ifnum\contacyrill=1     \def\cyrill{\eightwncyr}
  \def\cyrilb{\eightwncyb}  \def\cyrili{\eightwncyi}         \fi
  \textfont3=\eightex \scriptfont3=\sevenex \scriptscriptfont3=\sevenex
  \def\cmmib{\fam\cmmibfam\eightcmmib}  \textfont\cmmibfam=\eightcmmib
  \scriptfont\cmmibfam=\sixcmmib \scriptscriptfont\cmmibfam=\fivecmmib
  \def\cmbsy{\fam\cmbsyfam\eightcmbsy}  \textfont\cmbsyfam=\eightcmbsy
  \scriptfont\cmbsyfam=\sixcmbsy \scriptscriptfont\cmbsyfam=\fivecmbsy
  \def\cmcsc{\fam\cmcscfam\eightcmcsc} \scriptfont\cmcscfam=\eightcmcsc
  \textfont\cmcscfam=\eightcmcsc \scriptscriptfont\cmcscfam=\eightcmcsc
     \fi             \tt \ttglue=.5em plus.25em minus.15em
  \normalbaselineskip=9pt
  \setbox\strutbox=\hbox{\vrule height7pt depth2pt width0pt}
  \let\sc=\sixrm \let\big=\eightbig \normalbaselines\rm }
\gdef\tenbig#1{{\hbox{$\left#1\vbox to8.5pt{}\right.\n@space$}}}
\gdef\ninebig#1{{\hbox{$\textfont0=\tenrm\textfont2=\tensy
   \left#1\vbox to7.25pt{}\right.\n@space$}}}
\gdef\eightbig#1{{\hbox{$\textfont0=\ninerm\textfont2=\ninesy
   \left#1\vbox to6.5pt{}\right.\n@space$}}}
\def\alternativefont#1#2{\ifx\arisposta\amsrisposta \relax \else
\xdef#1{#2} \fi}
\global\contaeuler=0 \global\contacyrill=0 \global\contaams=0
%
%
%
%
\newbox\fotlinebb \newbox\hedlinebb \newbox\leftcolumn
\gdef\makeheadline{\vbox to 0pt{\vskip-22.5pt
     \fullline{\vbox to8.5pt{}\the\headline}\vss}\nointerlineskip}
\gdef\makehedlinebb{\vbox to 0pt{\vskip-22.5pt
     \fullline{\vbox to8.5pt{}\copy\hedlinebb\hfil
     \line{\hfill\the\headline\hfill}}\vss} \nointerlineskip}
\gdef\makefootline{\baselineskip=24pt \fullline{\the\footline}}
\gdef\makefotlinebb{\baselineskip=24pt
    \fullline{\copy\fotlinebb\hfil\line{\hfill\the\footline\hfill}}}
\gdef\doubleformat{\shipout\vbox{\Landspec\makehedlinebb
     \fullline{\box\leftcolumn\hfil\columnbox}\makefotlinebb}
     \advancepageno}
\gdef\columnbox{\leftline{\pagebody}}
\gdef\line#1{\hbox to\hsize{\hskip\leftskip#1\hskip\rightskip}}
\gdef\fullline#1{\hbox to\fullhsize{\hskip\leftskip{#1}%
\hskip\rightskip}}
\gdef\footnote#1{\let\@sf=\empty
         \ifhmode\edef\#sf{\spacefactor=\the\spacefactor}\/\fi
         #1\@sf\vfootnote{#1}}
\gdef\vfootnote#1{\insert\footins\bgroup
         \ifnum\dimnota=1  \eightpoint\fi
         \ifnum\dimnota=2  \ninepoint\fi
         \ifnum\dimnota=0  \tenpoint\fi
         \interlinepenalty=\interfootnotelinepenalty
         \splittopskip=\ht\strutbox
         \splitmaxdepth=\dp\strutbox \floatingpenalty=20000
         \leftskip=\oldssposta \rightskip=\olddsposta
         \spaceskip=0pt \xspaceskip=0pt
         \ifnum\sinnota=0   \textindent{#1}\fi
         \ifnum\sinnota=1   \item{#1}\fi
         \footstrut\futurelet\next\fo@t}
\gdef\fo@t{\ifcat\bgroup\noexpand\next \let\next\f@@t
             \else\let\next\f@t\fi \next}
\gdef\f@@t{\bgroup\aftergroup\@foot\let\next}
\gdef\f@t#1{#1\@foot} \gdef\@foot{\strut\egroup}
\gdef\footstrut{\vbox to\splittopskip{}}
\skip\footins=\bigskipamount
\count\footins=1000  \dimen\footins=8in
\catcode`@=12
\tenpoint
\ifnum\unoduecol=1 \hsize=\tothsize   \fullhsize=\tothsize \fi
\ifnum\unoduecol=2 \hsize=\collhsize  \fullhsize=\tothsize \fi
\global\let\lrcol=L      \ifnum\unoduecol=1
\output{\plainoutput{\ifnum\tipbnota=2 \clearnmbnota\fi}} \fi
\ifnum\unoduecol=2 \output{\if L\lrcol
     \global\setbox\leftcolumn=\columnbox
     \global\setbox\fotlinebb=\line{\hfill\the\footline\hfill}
     \global\setbox\hedlinebb=\line{\hfill\the\headline\hfill}
     \advancepageno  \global\let\lrcol=R
     \else  \doubleformat \global\let\lrcol=L \fi
     \ifnum\outputpenalty>-20000 \else\dosupereject\fi
     \ifnum\tipbnota=2\clearnmbnota\fi }\fi
\def\ifdoublepage{\ifnum\unoduecol=2 }
\gdef\yespagenumbers{\footline={\hss\tenrm\folio\hss}}
\gdef\ciao{ \ifnum\fdefcontre=1 \endfdef\fi
     \par\vfill\supereject \ifnum\unoduecol=2
     \if R\lrcol  \headline={}\nopagenumbers\null\vfill\eject
     \fi\fi \end}

\newskip\olddsposta \newskip\oldssposta
\global\oldssposta=\leftskip \global\olddsposta=\rightskip

\def\filldots{\leaders\hbox to 1em{\hss.\hss}\hfill}
\def\inquadrb#1 {\vbox {\hrule  \hbox{\vrule \vbox {\vskip .2cm
    \hbox {\ #1\ } \vskip .2cm } \vrule  }  \hrule} }
 \def\newline{\hfil\break}
\def\jump{\vskip\baselineskip} \newskip\iinnffrr
\def\sjump{\iinnffrr=\baselineskip
          \divide\iinnffrr by 2 \vskip\iinnffrr}
\def\bjump{\vskip\baselineskip \vskip\baselineskip}
\newcount\nmbnota  \def\clearnmbnota{\global\nmbnota=0}
\newcount\tipbnota \def\letterfootnote{\global\tipbnota=1}

\def\note#1{\global\advance\nmbnota by 1 \ifnum\tipbnota=1
    \footnote{$^{\rm\nttlett}$}{#1} \else {\ifnum\tipbnota=2
    \footnote{$^{\nttsymb}$}{#1}
    \else\footnote{$^{\the\nmbnota}$}{#1}\fi}\fi}
\def\nttlett{\ifcase\nmbnota \or a\or b\or c\or d\or e\or f\or
g\or h\or i\or j\or k\or l\or m\or n\or o\or p\or q\or r\or
s\or t\or u\or v\or w\or y\or x\or z\fi}
\def\nttsymb{\ifcase\nmbnota \or\dag\or\sharp\or\ddag\or\star\or
\natural\or\flat\or\clubsuit\or\diamondsuit\or\heartsuit
\or\spadesuit\fi}   \clearnmbnota
\def\numberfootnote{\global\tipbnota=0} \numberfootnote
\def\setnote#1{\expandafter\xdef\csname#1\endcsname{
\ifnum\tipbnota=1 {\rm\nttlett} \else {\ifnum\tipbnota=2
{\nttsymb} \else \the\nmbnota\fi}\fi} }
\newcount\nbmfig  \def\clearnbmfig{\global\nbmfig=0}
\gdef\figure{\global\advance\nbmfig by 1
      {\rm fig. \the\nbmfig}}   \clearnbmfig
\def\setfig#1{\expandafter\xdef\csname#1\endcsname{fig. \the\nbmfig}}
 \def\endformula{\eqno\numero $$}
 \def\efr{\endformula}
\newcount\frmcount \def\clearfrmcount{\global\frmcount=0}
\def\numero{\global\advance\frmcount by 1   \ifnum\indappcount=0
  {\ifnum\cpcount <1 {\hbox{\rm (\the\frmcount )}}  \else
  {\hbox{\rm (\the\cpcount .\the\frmcount )}} \fi}  \else
  {\hbox{\rm (\applett .\the\frmcount )}} \fi}
\def\nameformula#1{\global\advance\frmcount by 1%
\ifnum\draftnum=0  {\ifnum\indappcount=0%
{\ifnum\cpcount<1\xdef\spzzttrra{(\the\frmcount )}%
\else\xdef\spzzttrra{(\the\cpcount .\the\frmcount )}\fi}%
\else\xdef\spzzttrra{(\applett .\the\frmcount )}\fi}%
\else\xdef\spzzttrra{(#1)}\fi%
\expandafter\xdef\csname#1\endcsname{\spzzttrra}
\eqno \hbox{\rm\spzzttrra} $$}
\def\nfr{\nameformula}    \def\numali{\numero}
\def\nameali#1{\global\advance\frmcount by 1%
\ifnum\draftnum=0  {\ifnum\indappcount=0%
{\ifnum\cpcount<1\xdef\spzzttrra{(\the\frmcount )}%
\else\xdef\spzzttrra{(\the\cpcount .\the\frmcount )}\fi}%
\else\xdef\spzzttrra{(\applett .\the\frmcount )}\fi}%
\else\xdef\spzzttrra{(#1)}\fi%
\expandafter\xdef\csname#1\endcsname{\spzzttrra}
  \hbox{\rm\spzzttrra} }      \clearfrmcount
\newcount\cpcount \def\clearcpcount{\global\cpcount=0}
\newcount\subcpcount \def\clearsubcpcount{\global\subcpcount=0}
\newcount\appcount \def\clearappcount{\global\appcount=0}
\newcount\indappcount \def\clearindappcount{\indappcount=0}
\newcount\sottoparcount 

\def\applett{\ifcase\appcount  \or {A}\or {B}\or {C}\or
{D}\or {E}\or {F}\or {G}\or {H}\or {I}\or {J}\or {K}\or {L}\or
{M}\or {N}\or {O}\or {P}\or {Q}\or {R}\or {S}\or {T}\or {U}\or
{V}\or {W}\or {X}\or {Y}\or {Z}\fi    \ifnum\appcount<0
\immediate\write16 {Panda ERROR - Appendix: counter "appcount"
out of range}\fi  \ifnum\appcount>26  \immediate\write16 {Panda
ERROR - Appendix: counter "appcount" out of range}\fi}
\clearappcount  \clearindappcount \newcount\connttrre
\def\clearconnttrre{\global\connttrre=0} \newcount\countref
\def\clearcountref{\global\countref=0} \clearcountref
\def\chapter#1{\global\advance\cpcount by 1 \clearfrmcount
                 \goodbreak\null\vbox{\jump\nobreak
                 \clearsubcpcount\clearindappcount
                 \itemitem{\ttaarr\the\cpcount .\qquad}{\ttaarr #1}
                 \par\nobreak\jump\sjump}\nobreak}
\def\section#1{\global\advance\subcpcount by 1 \goodbreak\null
               \vbox{\sjump\nobreak\ifnum\indappcount=0
                 {\ifnum\cpcount=0 {\itemitem{\ppaarr
               .\the\subcpcount\quad\enskip\ }{\ppaarr #1}\par} \else
                 {\itemitem{\ppaarr\the\cpcount .\the\subcpcount\quad
                  \enskip\ }{\ppaarr #1} \par}  \fi}
                \else{\itemitem{\ppaarr\applett .\the\subcpcount\quad
                 \enskip\ }{\ppaarr #1}\par}\fi\nobreak\jump}\nobreak}
\clearsubcpcount
\def\appendix#1{\global\advance\appcount by 1 \clearfrmcount
                  \goodbreak\null\vbox{\jump\nobreak
                  \global\advance\indappcount by 1 \clearsubcpcount
          \itemitem{ }{\hskip-40pt\ttaarr Appendix\ \applett :\ #1}
             \nobreak\jump\sjump}\nobreak}
\clearappcount \clearindappcount
\def\references{\goodbreak\null\vbox{\jump\nobreak
   \itemitem{}{\ttaarr References} \nobreak\jump\sjump}\nobreak}

\def\introduction{\clearindappcount\clearappcount\clearcpcount
                  \clearsubcpcount\goodbreak\null\vbox{\jump\nobreak
  \itemitem{}{\ttaarr Introduction} \nobreak\jump\sjump}\nobreak}
\clearcpcount\clearcountref
\def\acknowledgements{\goodbreak\null\vbox{\jump\nobreak
\itemitem{ }{\ttaarr Acknowledgements} \nobreak\jump\sjump}\nobreak}
\def\setchap#1{\ifnum\indappcount=0{\ifnum\subcpcount=0%
\xdef\spzzttrra{\the\cpcount}%
\else\xdef\spzzttrra{\the\cpcount .\the\subcpcount}\fi}
\else{\ifnum\subcpcount=0 \xdef\spzzttrra{\applett}%
\else\xdef\spzzttrra{\applett .\the\subcpcount}\fi}\fi
\expandafter\xdef\csname#1\endcsname{\spzzttrra}}
\newcount\draftnum \newcount\ppora   \newcount\ppminuti
\global\ppora=\time   \global\ppminuti=\time
\global\divide\ppora by 60  \draftnum=\ppora
\multiply\draftnum by 60    \global\advance\ppminuti by -\draftnum
\def\droggi{\number\day /\number\month /\number\year\ \the\ppora
:\the\ppminuti}     \global\draftnum=0
\def\draftcomment#1{\ifnum\draftnum=0 \relax \else
{\ {\bf ***}\ #1\ {\bf ***}\ }\fi} 
%
%
\catcode`@=11
\gdef\Ref#1{\expandafter\ifx\csname @rrxx@#1\endcsname\relax%
{\global\advance\countref by 1    \ifnum\countref>200
\immediate\write16 {Panda ERROR - Ref: maximum number of references
exceeded}  \expandafter\xdef\csname @rrxx@#1\endcsname{0}\else
\expandafter\xdef\csname @rrxx@#1\endcsname{\the\countref}\fi}\fi
\ifnum\draftnum=0 \csname @rrxx@#1\endcsname \else#1\fi}
\gdef\beginref{\ifnum\draftnum=0  \gdef\Rref{\fairef}
\gdef\endref{\scriviref} \else\relax\fi
\ifx\risposta\mplarisposta \ninepoint \fi
\baselineskip=12pt \parskip 2pt plus.2pt }
\def\Reflab#1{[#1]} \gdef\Rref#1#2{\item{\Reflab{#1}}{#2}}
\gdef\endref{\relax}  \newcount\conttemp
\gdef\fairef#1#2{\expandafter\ifx\csname @rrxx@#1\endcsname\relax
{\global\conttemp=0 \immediate\write16 {Panda ERROR - Ref: reference
[#1] undefined}} \else
{\global\conttemp=\csname @rrxx@#1\endcsname } \fi
\global\advance\conttemp by 50  \global\setbox\conttemp=\hbox{#2} }
\gdef\scriviref{\clearconnttrre\conttemp=50
\loop\ifnum\connttrre<\countref \advance\conttemp by 1
\advance\connttrre by 1
\item{\Reflab{\the\connttrre}}{\unhcopy\conttemp} \repeat}
\clearcountref \clearconnttrre
\catcode`@=12
\ifx\risposta\mplarisposta \def\Reflab#1{#1.} \letterfootnote \fi

\def\slashchar#1{\setbox0=\hbox{$#1$} \dimen0=\wd0
     \setbox1=\hbox{/} \dimen1=\wd1 \ifdim\dimen0>\dimen1
      \rlap{\hbox to \dimen0{\hfil/\hfil}} #1 \else
      \rlap{\hbox to \dimen1{\hfil$#1$\hfil}} / \fi}
\ifx\oldchi\undefined \let\oldchi=\chi
  \def\cchi{{\raise 1pt\hbox{$\oldchi$}}} \let\chi=\cchi \fi
  
\def\del{\partial}   

\def\frac#1#2{{\textstyle{#1 \over #2}}}

\def\half{\ifinner {\scriptstyle {1 \over 2}}\else {1 \over 2} \fi}
\def\bra#1{\langle#1\vert}  \def\ket#1{\vert#1\rangle}

\def\vev#1{\langle#1\rangle}
\def\bivev#1#2{\langle#1\vert#2\rangle}
\def\simge{\rlap{\raise 2pt \hbox{$>$}}{\lower 2pt \hbox{$\sim$}}}
\def\simle{\rlap{\raise 2pt \hbox{$<$}}{\lower 2pt \hbox{$\sim$}}}

\def\buildchar#1#2#3{{\null\!\mathop{#1}\limits^{#2}_{#3}\!\null}}

\def\vbig#1#2{{\vbigd@men=#2\divide\vbigd@men by 2%
\hbox{$\left#1\vbox to \vbigd@men{}\right.\n@space$}}}

\def\noblackbox{\overfullrule=0pt} \def\yesblackbox{\overfullrule=5pt}
%
%
\newcount\fdefcontre \newcount\fdefcount \newcount\indcount
\newread\filefdef  \newread\fileftmp  \newwrite\filefdef
\newwrite\fileftmp     \def\strip#1*.A {#1}
\def\futuredef#1{\beginfdef
\expandafter\ifx\csname#1\endcsname\relax%
{\immediate\write\fileftmp {#1*.A}
\immediate\write16 {Panda Warning - fdef: macro "#1" on page
\the\pageno \space undefined}
\ifnum\draftnum=0 \expandafter\xdef\csname#1\endcsname{(?)}
\else \expandafter\xdef\csname#1\endcsname{(#1)} \fi
\global\advance\fdefcount by 1}\fi   \csname#1\endcsname}

\def\beginfdef{\ifnum\fdefcontre=0
\immediate\openin\filefdef \jobname.fdef
\immediate\openout\fileftmp \jobname.ftmp
\global\fdefcontre=1  \ifeof\filefdef \immediate\write16 {Panda
WARNING - fdef: file \jobname.fdef not found, run TeX again}
\else \immediate\read\filefdef to\spzzttrra
\global\advance\fdefcount by \spzzttrra
\indcount=0      \loop\ifnum\indcount<\fdefcount
\advance\indcount by 1   \immediate\read\filefdef to\spezttrra
\immediate\read\filefdef to\sppzttrra
\edef\spzzttrra{\expandafter\strip\spezttrra}
\immediate\write\fileftmp {\spzzttrra *.A}
\expandafter\xdef\csname\spzzttrra\endcsname{\sppzttrra}
\repeat \fi \immediate\closein\filefdef \fi}
\def\endfdef{\immediate\closeout\fileftmp   \ifnum\fdefcount>0
\immediate\openin\fileftmp \jobname.ftmp
\immediate\openout\filefdef \jobname.fdef
\immediate\write\filefdef {\the\fdefcount}   \indcount=0
\loop\ifnum\indcount<\fdefcount    \advance\indcount by 1
\immediate\read\fileftmp to\spezttrra
\edef\spzzttrra{\expandafter\strip\spezttrra}
\immediate\write\filefdef{\spzzttrra *.A}
\edef\spezttrra{\string{\csname\spzzttrra\endcsname\string}}
\iwritel\filefdef{\spezttrra}
\repeat  \immediate\closein\fileftmp \immediate\closeout\filefdef
\immediate\write16 {Panda Warning - fdef: Label(s) may have changed,
re-run TeX to get them right}\fi}
\def\iwritel#1#2{\newlinechar=-1
{\newlinechar=`\ \immediate\write#1{#2}}\newlinechar=-1}
\global\fdefcontre=0 \global\fdefcount=0 \global\indcount=0
%
%
\null
%
%
%
%
%
\loadamsmath\chapterfont{\bf} \sectionfont{\scaps}
\def\sssty{\scriptscriptstyle}   \def\v{{\cal V}}
\def\Reflab#1{#1.}
\letterfootnote
\nopagenumbers
{\baselineskip=12pt
\line{\hfill PUPT-1348}
\line{\hfill hep-th/9210105}
\line{\hfill October, 1992}}
{\baselineskip=14pt
\ifdoublepage \bjump\bjump\else\bjump\bjump\jump\fi
\centerline{\capsone INFINITE SYMMETRY AND WARD IDENTITIES}
\sjump
\centerline{\capsone IN TWO-DIMENSIONAL STRING THEORY
\footnote{$^\dagger$}{Lectures delivered by I.R. Klebanov at
the Workshop ``String Quantum Gravity and Physics at the Planck Energy
Scale", Erice, June 21--28, 1992, and at the 1992 Trieste Summer School
of Theoretical Physics.} }
\ifdoublepage \bjump \jump \else \bjump \bjump \fi
\centerline{\scaps IGOR R. KLEBANOV and ANDREA PASQUINUCCI}
\sjump
\centerline{\it Joseph Henry Laboratories, Department of Physics,}
\centerline{\it Princeton University, Princeton, NJ 08544, USA}
\vfill
\centerline{\rm ABSTRACT}
\sjump
\noindent We review some of the recent progress
in the continuum formulation of
two-dimensional string theory, i.e. two-dimensional quantum gravity
coupled to $c=1$ matter. Special attention is devoted to the
discrete states and to the $w_\infty$
algebra they generate. To demonstrate the power of the infinite
symmetry, we use the $w_\infty$ Ward identities to derive
recursion relations among certain classes of correlation
functions, which allow to calculate them exactly.
\jump
\ifdoublepage\jump\else\bjump\jump\fi
\pageno=0 \eject \null \ifdoublepage\bjump\fi}
\yespagenumbers\pageno=1
\introduction
\noindent The field of two-dimensional quantum gravity has experienced
remarkable progress in recent years. Application of matrix model
techniques has led to exact solutions of various $c\leq 1$
conformal field theories coupled to
gravity.~$^{\Ref{mxmod},\Ref{GM},\Ref{Doug},\Ref{GMil},\Ref{GKleb}}$
Since string theory in the
first-quantized formulation reduces to 2-d quantum gravity, these
low-dimensional models should provide us with new insights into
``stringy'' phenomena. In particular, the $c=1$ model corresponds to
bosonic strings in {\it two} dimensions, the extra dimension originating
from the world sheet conformal factor.~$^{\Ref{Polch}}$

We have learned that, after coupling to 2-d gravity, theories simplify
considerably and become fully integrable. A good understanding of
this exact integrability is still missing. With this goal in mind,
it is important to develop the continuum path integral formulation
of 2-d gravity,~$^{\Ref{Liouv},\Ref{ddk}}$ which so far has lagged behind
the matrix models in its power. In these notes we will focus on
the $c=1$ model, i.e. quantum gravity coupled to one scalar
field.~$^{\Ref{GMil},\Ref{GKleb}}$
This model is the richest among those exactly
solved, and is also one of the simplest. Since there are several
reviews documenting the recent progress in the $c=1$ matrix
models,~$^{\Ref{I}, \Ref{vk}}$ we will discuss only the continuum
formulation where some new insights have recently been obtained.

After reviewing the basics of the continuum path integral approach,
we will proceed to the derivation of the $w_\infty$ symmetry
structure.~$^{\Ref{W},\Ref{KP},\Ref{WZ}}$
A similar $w_\infty$ symmetry has also appeared in the matrix model
approach,~$^{\Ref{aj}}$ but a precise connection between
the two is still missing. We will review the construction of
the ground ring operators,~$^{\Ref{W}}$ and of the $w_\infty$
currents.
As an explicit application of the symmetries, we will use the Ward
identities to calculate a large class of correlation
functions.~$^{\Ref{K},\Ref{KPas}}$
\chapter{Continuum description of 2D quantum gravity
coupled to c=1 matter}
\noindent In this section we give a brief review of the continuum
formulation of 2-dimensional string theory. We start by recalling
the path integral approach to non-critical
string theory in $D=c+1$ dimensions, and then apply it to
the $c=1$ case.
\section{Non-critical string theory in $c$ dimensions}
\noindent In the Polyakov~$^{\Ref{Liouv}}$ approach,
first-quantized
strings propagating in ${\Bbb R}^c$ are described as a theory of $c$
free bosons coupled to 2-d quantum gravity. In other words, we begin
with the path integral for
2-d quantum gravity coupled to $c$ scalar fields $X^i
(\buildchar{\sigma}{\rightarrow}{ })$~\note{This partition function
can be seen as a sum over random surfaces (2D geometries) embedded
in $c$ dimensions.}
$$\eqalignno{
&{\cal Z}\ =\ {1\over {\rm Vol}\left({\rm Diff}\right) }
 \int\left[Dg_{\mu\nu}(\buildchar{\sigma}{\rightarrow}{ })
\right]\otimes \prod_{i=1}^c \left[D
X^i(\buildchar{\sigma}{\rightarrow}{ })\right] \ e^{-I(g_{\mu\nu},
X^i)} \cr
&I(g_{\mu\nu}, X^i)\ =\ \frac1{8\pi}\int d^2\sigma \sqrt{g} \left(
g^{\mu\nu} \del_\mu X^i \del_\nu X^i + \lambda \right)\ . &\numali\cr}
$$
Renewed interest in this problem was stimulated in part by the
remarkable progress of the discretized (matrix model) approach, where
${\cal Z}$ was calculated for
$c\leq 1$.~$^{\Ref{mxmod}-\Ref{GKleb}}$
To study the problem in the continuum approach, it is convenient to
choose the conformal gauge~$^{\Ref{Liouv}}$
$$
g_{\mu\nu} (\buildchar{\sigma}{\rightarrow}{ })\ =\ e^{-\gamma \phi
(\buildchar{\sigma}{\rightarrow}{ })} \hat{g}_{\mu\nu}
(\buildchar{\sigma}{\rightarrow}{ },\tau)
\efr
where $\hat{g}_{\mu\nu}(\buildchar{\sigma}{\rightarrow}{ },\tau)$ is a
family of reference metrics parametrized by the moduli space.
Classically the Liouville field
$\phi(\buildchar{\sigma}{\rightarrow}{ })$ is non-dynamical, but in
the quantum case there is a contribution from the path integral measure
which gives rise to the kinetic term for it.$^{\Ref{Liouv}}$ ~\note{The
interested reader is referred to the original
literature~$^{\Ref{Gervais},\Ref{ddk},\Ref{dhoker}}$ for a
more detailed derivation of the following results. For a review,
see Ref. \Ref{ns}.} After gauge fixing, the integration measure
becomes
$$
\left[Dg\right]\otimes\left[DX^i\right]_g \ =\ \left[D({\rm Diff})
\right]
\otimes \left[d\tau\right] \otimes \left[D\phi\right]_g \otimes
\left[Db\, Dc\right]_g\otimes \left[ DX^i\right]_g
e^{-I(\hat g, b, c)}
\efr
where $b$ and $c$ are the ghosts, and $I$ is the standard ghost action.
$[d\tau]$ is the measure for integration over the moduli
which will not interest us since
eventually we will focus on the genus zero case.

Under a Weyl rescaling $g \rightarrow e^\psi g$, the matter and ghost
actions are invariant, whereas the measures change  according to
$$
\left[DX\right]_{e^\psi g}\otimes \left[Db\, Dc\right]_{e^\psi g} \ =\
 e^{ \left({ c-26\over 48\pi}\right) S_L(\psi,g)}  \left[DX\right]_g
\otimes \left[Db\, Dc\right]_g
\nfr{weylr}
and the Liouville action is given by
$$
S_L(\psi,g)\ =\ \int d^2 \sigma \sqrt{g} \left( \frac12 g^{\mu\nu}
\del_\mu\psi \del_\nu\psi + R\psi + \mu e^\psi\right)\ .
\efr
Since we have to integrate over $\phi$, we would like to change
from the measure defined with the field-dependent metric $g$,
to that defined with the fiducial metric $\hat g$. While the
transformation of the matter and ghost measures is simply given by
Eq. \weylr, the treatment of $[D\phi]$ is more subtle. According to a
conjecture of David, Distler and Kawai,~$^{\Ref{ddk}}$ later confirmed
in Ref. \Ref{dhoker}, the Jacobian for the change of measure from
$\left[D\phi\right]_g$ to
$\left[D\phi\right]_{\hat{g}}$
is given by the exponential of a renormalizable local action consistent
with the general coordinate invariance of the underlying theory.
Thus, the partition function on the sphere becomes
$$\eqalignno{
&{\cal Z}\ =\ \int \left[D\phi\right]_{\hat{g}} \otimes
\left[Db\, Dc\right]_{\hat{g}}\otimes
\prod_{i=1}^c \left[ DX^i\right]_{\hat{g}} e^{-I'(\hat{g}, X,\phi)
-  I(\hat{g},b,c)}&\numali\cr
&I'(\hat{g}, X,\phi)\ =\ \frac1{8\pi} \int d^2 \sigma \sqrt{\hat{g}}
\left[\hat{g}^{\mu\nu} \left(\del_\mu X^i\del_\nu X^i + \del_\mu \phi
\del_\nu \phi \right) - Q \hat{R} \phi + \Delta e^{\alpha\phi}
\right]\ . \cr}
$$
where the coefficients $Q$ and $\alpha$ are to be fixed by
quantum Weyl invariance with respect to $\hat g$,
which is the remnant of the general covariance
in the conformal gauge.
We can now view this as a sigma model for
ordinary string theory in flat $D=c+1$
dimensions with a dilaton condensate $\Phi = -Q\phi$, and a ``tachyon"
condensate $T = \Delta e^{\alpha\phi}$.

To fix $Q$ and $\alpha$, one uses
the requirement of cancellation of the conformal anomaly and the
fact that a physical operator must have dimension $(1,1)$.
First setting $\Delta =0$, the chiral stress-energy tensor is
$$\eqalignno{
&T^{(X,\phi)}_{zz}\ =\ -\frac12 (\del_z X^i)^2 -\frac12 (\del_z\phi)^2
- \frac12 Q \del^2_z \phi \cr
&T^{(b,c)}_{zz}\ =\ -2 b_{zz} \del_z c^z + c^z \del_z b_{zz} \ .
&\numali\cr}
$$
The Fourier components $L_n = \frac1{2\pi i}
\oint dz\, z^{n+1} T^{(X,\phi)}_{zz}$ form the Virasoro algebra
$$
[L_n,L_m]\ =\ (n-m)L_{n+m} + {c+1 +3Q^2\over 12} \, n(n^2-1)\,
\delta_{n+m,0}\ .
\efr
To cancel the ghost contribution  to the conformal anomaly one must
set $c+1 + 3 Q^2 - 26 =0$, i.e.
$$
Q\ =\ \sqrt{{25-c\over 3}}\ .
\efr
Further, by requiring $e^{\alpha\phi}$ to be a dimension $(1,1)$
perturbation, one finds~$^{\Ref{ddk}}$
$$
\alpha \ =\ - {1\over 2\sqrt3}\left(\sqrt{25-c} - \sqrt{1-c}\right)\ .
\efr
For $c>1$ we find the problem of complex $\alpha$, which is indicative
of the tachyonic nature of the string theory. Instead, we will set $c$
to its critical value $c=1$.
\section{2D quantum gravity coupled to $c=1$ matter}
\noindent
Applying the previous formalism, we obtain
$Q=2\sqrt2$, $\alpha=-\sqrt2$.
The 2-d quantum gravity coupled to $c=1$
matter can be viewed as a string theory in flat $D=c+1=2$ dimensions
with a dilaton condensate and a tachyon condensate (if $\Delta \neq
0$). This is the highest number of dimensions where bosonic strings
are tachyon-free. We will see that the would-be tachyon of $D>2$ theory
is exactly massless for $D=2$. The reader may be surprised by the
lack of Poincar\'e invariance, since the
two spacetime coordinates ($X,\phi$) appear on
a different footing.~\note{The lack of
translation invariance was also
found in the matrix model formulation of 2-d
gravity coupled to $c=1$ matter.~$^{\Ref{dj}}$}
Surprisingly, the theory instead possesses an infinite hidden symmetry.
The main purpose of these notes is to make this infinite symmetry
explicit.

Let us start by looking for the simplest physical states in this theory.
We know that a physical state $\ket{\psi}$ must satisfy
$$\eqalignno{
& L_n\ket{\psi}\ =\ \overline{L}_n \ket{\psi}\ =\ 0 \qquad\qquad {\rm
for}\ n>0\ \cr
& L_0\ket{\psi}\ =\ \overline{L}_0 \ket{\psi}\ =\ 1\,\cdot\,
\ket{\psi}\ . &\nameali{physcond}\cr}
$$
The physical states are defined modulo pure gauge states, which are
the Virasoro descendants.
The simplest physical states are those with no oscillator excitations,
characterizing motion of a string in its ground state,
$$
\ket{p,\epsilon}\ =\ e^{ipX+\epsilon\phi}(0)\ket{0}\ .
\efr
They satisfy $L_n \ket{p,\epsilon} = \overline{L}_n \ket{p,\epsilon} =
0$, $n>0$,  and
$$
L_0 \ket{p,\epsilon} = \overline{L}_0 \ket{p,\epsilon}\ = \ \left[
\frac12 p^2 -\frac12 \epsilon(\epsilon+2\sqrt2)\right]
\ket{p,\epsilon}\ .
\efr
Then the on-shell conditions Eqs. \physcond\ imply $p^2 -
\epsilon(\epsilon+2\sqrt2)=2$ or, defining $E=\epsilon +\sqrt2$,
$$
p^2 -E^2\ =\ 0\ .
\efr
This is a massless dispersion relation. In other words, the ``tachyon"
is massless in two dimensions, in accordance with the
usual formula $m_T^2=(2-D)/12$. The solution of the dispersion
relation is
$\epsilon = -\sqrt2 +\chi p$, where $\chi = \pm 1$ is the chirality.
The associated tachyon vertex operators are
$$
T_\chi (p)\ =\ \int d^2\sigma \sqrt{\hat{g}}\, e^{ipX +(\chi p
-\sqrt2)\phi}=
\int d^2\sigma \sqrt{\hat{g}}\, g_{st} (\phi) e^{ipX +\chi p \phi}
\ .
\efr
The factor $e^{-\sqrt2 \phi}=g_{st}(\phi)$ is the position-dependent
string coupling constant.~\note{We may formally continue to
Minkowski signature by sending $\phi \rightarrow i\, t$
and interpreting $t$ as the time. In such a theory, $T_\chi$ correspond
to the ordinary plane waves. The unusual feature is that
the string coupling
$g_{st}(t)=e^{-i\sqrt2 t}$ is complex and oscillatory.}
Setting $p=0$ we find again $\alpha=-\sqrt2$.
The operator $e^{-\sqrt2 \phi}$ is believed to be exactly marginal.

We will be interested in the correlation
functions of tachyons (on the sphere)
$$
\vev{\prod_{i=1}^N T_{\chi_i}(p_i) }\ .
\nfr{tacvev}
Since the theory is translationally
invariant in $X$, the correlator is non-vanishing only if
$\sum_i p_i =0$. On the other hand, $\epsilon$ is not
conserved. There is, however, a class of correlation functions where
the integrand does not depend on the zero mode $\phi_0$ of $\phi$ in
the $\Delta \rightarrow 0$ limit. The $\phi_0$
dependence of the integrand in Eq. \tacvev\ is then given by
$$
{\rm Exp}\left[\sum_{i=1}^N \epsilon_i \phi_0 + \frac1{8\pi} \int d^2
\sigma \sqrt{\hat{g}} Q \hat{R} \phi_0 \right]\ =\ {\rm Exp} \left[
\left( \sum_{i=1}^N \epsilon_i + 2\sqrt2 \right) \phi_0 \right]\ .
\efr
We will focus on the correlators which satisfy the sum rule
$$
\sum_{i=1}^N \epsilon_i\ =\ -2\sqrt2
\nfr{srule}
analogous to energy conservation. These correlators are sometimes called
resonant because they are enhanced by the volume of $\phi_0$.~\note{Note
that in the $\phi\to it$ continued theory, Eq. \srule\ is enforced
by a delta-function constraint.}
Defining $X = X_0 + \widetilde{X}$ where $X_0$ is the zero mode of
$X$, and similarly for $\phi$,
we can explicitly perform the zero-mode integrals. Then the resonant
correlators become~$^{\Ref{Sasha},\Ref{GK}}$
$$
\vev{\prod_{i=1}^N T_{\chi_i}(p_i) }\ =\ \frac1{\sqrt2}\, \vert \log
\Delta \vert \ \delta (\sum_{i=1}^N p_i)\  A(\chi_i, p_i)
\efr
where
$$\eqalignno{
& A(\chi_i, p_i)\ =\ \int [D\widetilde{X}] [D\widetilde{\phi}] \left(
\prod_{i=1}^N \int d^2 \sigma \sqrt{\hat{g}} e^{ip_i \widetilde{X} +
\epsilon_i \widetilde\phi}\right) e^{-S(\widetilde{X},
\widetilde\phi)}\cr
& S(\widetilde{X}, \widetilde\phi)\ =\ \frac1{8\pi} \int d^2\sigma
\sqrt{\hat{g}} \left( \del_\mu \widetilde{X}\del^\mu \widetilde{X} +
\del_\mu \widetilde\phi \del^\mu \widetilde\phi \right)\ . &\numali
\cr} $$
The amplitudes $A(\chi_i, p_i)$ are sometimes
called the ``{\sl Shifted Virasoro-Shapiro\/}"
amplitudes.~$^{\Ref{Sasha},\Ref{GK}}$
Since they are formulated in terms of free fields $\widetilde{X}$,
$\widetilde\phi$, they can be written down explicitly and checked to
be $SL(2, C)$ invariant. After fixing the positions of three vertex
operators, we find
$$
A_{\chi_1, \dots, \chi_N} (p_1, \dots , p_N)\ =\ \int
\prod_{i=1}^{N-3} d^2 z_i\, \prod _{i<j} \vert z_i - z_j \vert^{2 f_i
\cdot f_j}
\efr
where $f_i \cdot f_j = p_i p_j - \epsilon_i \epsilon_j$ (and
$\epsilon_i = - \sqrt2 + \chi_i p_i$).

These amplitudes exhibit a remarkable structure. They do not
vanish only if there is exactly one particle with $\chi = -1$
(or with $\chi= +1
$).~$^{\Ref{Sasha},\Ref{GK},\Ref{kdf}}$
Then one finds
$$
A_{+ \dots +-}^{(N,1)} (p_1, \dots , p_N, p_{N+1})\ =\ {\pi^{N-2}
\over (N-2)! } \prod_{i=1}^N {\Gamma(1 -\sqrt2 p_i )\over
\Gamma(\sqrt2 p_i)}
\nfr{ampli}
for amplitudes of type $(N,1)$, i.e. for $N$ particles with
$\chi = +1$ and one particle with $\chi=-1$.
The sum rules $\sum_i p_i =0$ and $\sum_i
\epsilon_i = -2\sqrt2$ fix, in this case, the value of $p_{N+1}$.
Indeed, we have
$$
\sum_{i=1}^N p_i \ =\ - p_{N+1}\ =\ \frac1{\sqrt2} (N-1)\ .
\efr
The amplitudes of type $(1, N)$ are obtained from Eq. \ampli\ by a
parity flip.
\section{The appearance of the ``Discrete States"}
\noindent The surprising feature of the amplitudes Eq. \ampli\  is the
presence of poles at $p_i=n/\sqrt2$ for each $\chi=1$ external leg.
The infinite
sequence of poles is related to the existence of operators other than
the ``tachyons".~\note{These states were first found in the matrix
model formulation of 2-d quantum gravity.~$^{\Ref{GKN}}$}
Indeed, consider the four-point function of type $(3, 1)$,
where $p_4=-\sqrt2$ is fixed by the sum rules.
The poles result from the integration
over the coordinates of $T_-(p_4)$ near one of the other vertex
operators. Consider the region where $T_-(p_4)$ approaches
$T_+(p_1)$. Let us examine the origin of the first pole which
occurs at $p_1=1/\sqrt2$. The relevant operator product expansion
({\scaps o.p.e.}) is
$$\eqalignno{
p_1=\frac1{\sqrt2}\,:\qquad T_-(-\sqrt2)& \, \cdot \,
T_+  (\frac1{\sqrt2})
\ =\  e^{-i\sqrt2 X}(z,\bar{z}) \cdot e^{\frac{i}{\sqrt2} X -
\frac1{\sqrt2}\phi}(0) \ \sim\ \cr
& {1\over \vert z\vert^2} e^{-\frac{i}{\sqrt2}
X - \frac1{\sqrt2}\phi}(0) \ =\ {1\over \vert z\vert^2}
T_-(-\frac1{\sqrt2})\ .&\numali\cr}
$$
After integrating over $z$, we find a divergence due to the appearance
of the on-shell ``tachyon'' $T_-(-\frac1{\sqrt2})$
in the intermediate channel. This only explains the origin of the first
pole in $p_1$, however. Can an infinite sequence of Regge poles occur
in a theory of a single scalar field? The answer is obviously negative.
Indeed, repeating the calculation for $p_1=\sqrt2$, we find that
the operator appearing in the intermediate channel contains
oscillator excitations,
$$
p_1=\sqrt2\,:\qquad\quad  e^{-i\sqrt2 X}(z,\bar{z}) \cdot e^{i\sqrt2 X}
(0) \ \sim\ {1\over \vert z\vert^2} \del X\overline\del X\ .
\efr
$\del X\overline\del X$ is a new operator
called the ``{\sl dilaton\/}", which can easily be checked to
satisfy the Virasoro conditions. In fact, at each subsequent pole we
find a new physical operator. The main feature of all
these operators is that they are physical only at special values of
the momenta and do not give rise to propagating states.

Let us look for such ``{\sl discrete\/}" states more systematically
by constructing physical states (which satisfy Eq.
\physcond) using the oscillators of the $X$ and $\phi$ fields.
Introducing the oscillators $\alpha_n$ and $\beta_n$ through
$$
\del_z X\ =\ -i \sum_n \alpha_n z^{-n-1}\ , \qquad\quad \del_z \phi\
=\ -i \sum_n \beta_n z^{-n-1}
\efr
we define
$$
\alpha^\mu_n\ =\ (i\beta_n, \alpha_n)\ , \qquad f^\mu\ =\
(\epsilon,p)\ ,\qquad Q^\mu\ =\ (Q,0) \ .
\efr
The Virasoro generators take the form
$$\eqalignno{
& L_n\ =\ (f^\mu +\frac{n+1}2 Q^\mu) \alpha_{\mu,n} + \sum_k :
\alpha^\mu_{n+k} \alpha_{\mu, -k} : \qquad \quad n\neq 0 \cr
& L_0 \ =\ f_\mu ( f^\mu + Q^\mu) + \sum_k : \alpha^\mu_{-k}
\alpha_{\mu,k} : &\numali\cr}
$$
where $[\alpha^\nu_n , \alpha^\mu_m] = n\,\delta_{n+m,0}\,
\eta^{\mu\nu}$, $\eta_{\mu\nu} = {\rm diag}(-1,1)$ and the
indices are raised and lowered with the $\eta$ metric.

We have already seen that the ``tachyons'' are the physical states
without any oscillator excitations. Consider now the states at level
one, of the form
$$
\ket{\psi}\ =\ e_{\mu\nu}\, \alpha^\mu_{-1}\overline{\alpha}^\nu_{-1}\,
\ket{f}\ .
\efr
We simplify the discussion by considering only the chiral half of this
state (relevant to the open string case) where
$$
\ket{\psi_L}\ =\ e_{\mu}\, \alpha^\mu_{-1}\, \ket{f}\ .
\efr
We can later construct closed-string states by setting $e_{\mu\nu} =
e_\mu e_\nu$.~\note{Obviously it is not guaranteed that this
procedure exhausts all the possible physical closed-string states.
We will see an example of this in section \S 3.2.}

The Virasoro conditions, Eq. \physcond,  now give
$$
(f_\mu + Q_\mu)\, e^\mu(f)\ =\ 0 \ , \qquad\quad
(f_\mu + Q_\mu)\, f^\mu\ =\ 0 \ .
\efr
Notice also that the state $f_\mu \alpha^\mu_{-1} \ket{f}$ (with
polarization $e^\mu = f^\mu$) is a pure gauge state, $L_{-1} \ket{f}$.

For a general $f_\mu$, the unique solution of the Virasoro conditions
is $e_\mu \sim f_\mu$, which corresponds to the pure gauge state. This
is in accord with the na\"\i ve light-cone argument that there are no
physical oscillator states in two-dimensional string theory.

\noindent There are two exceptional momenta, however.

\item{$\bullet$}{$f^\mu = 0$. ~ Here the gauge
symmetry becomes trivial,
and the polarization $e_\mu=(0,1)$ gives rise to a non-trivial
physical state, whose vertex operator is
$\del X\overline\del X$.}
\item{$\bullet$}{$ f^\mu =- Q^\mu$. ~ In this case the constraints
are trivially satisfied, and the polarization $e_\mu = (0,1)$ again
gives a physical state. The corresponding operator is $\del
X\overline\del X e^{-2\sqrt2 \phi}$.}

\noindent Notice that the new vertex operators are
two different Liouville ``dressings'' of the pure matter primary field
$\del X\overline\del X$.

Continuing this analysis to higher levels, one can
construct a family of physical operators~$^{\Ref{KP}}$
which are matter ($X$)
primary fields appropriately dressed by exponentials of
the Liouville field $\phi$. These operators are exhibited in
section \S 3.1. It was proven in Refs. \Ref{lz}, \Ref{Bouw} that these
are all the physical states not containing ghost excitations.
However, there is also an infinite set of ``ghostly'' operators
that turn out to play an important physical r\^ole in this theory.
We will discuss them in the next section, after introducing the
BRST formalism as a necessary tool.
\chapter{Discrete states and the ``Ground Ring"}
\noindent First we change our conventions rescaling
$\alpha'$ so as to eliminate the factors $\sqrt2$ from our
formulas. Setting $\alpha' = 4$ (instead of $\alpha' = 2$)
one has $Q=2$, and the special states occur at momenta $p=n/2$.
With these conventions $\vev{X(z,\bar{z}) X(w,\bar{w})}=-2 \log \vert
z-w\vert^2$, and the tachyon operators are given by
$$\eqalignno{& T^\pm_k(z,\bar{z})\ =
\ e^{ikX+\epsilon_\pm\phi}(z,\bar{z})\cr
&\epsilon_\pm=-1\pm k&\numali\cr}
$$
The sum rules for the resonant tachyon correlators become
$$
\sum_{i=1}^n k_i\ =\ 0\ , \qquad\quad \sum_{i=1}^n \epsilon_i\ =\ -2\ .
\nfr{sumrules}
We will keep these conventions for the rest of the paper.
\section{``Discrete States" and the $c=1$ primary
fields}
\noindent As we concluded in section \S 2.3, some of the discrete states
of the two-dimensional string theory are simply primary fields of the
$c=1$ CFT dressed by exponentials of the Liouville field.
\note{The presence of special discrete primary fields in the $c=1$ CFT
has been known for a long time.~$^{\Ref{discrete}}$}
To construct
all such states, we review the structure of the $c=1$ primary fields.
First we consider the theory compactified at
the self-dual radius $R=2$,
which is well known to have a chiral $SU(2)$ current algebra
generated by
$$
H_\pm(z)\ =\ \oint {du\over 2\pi i} : e^{\pm i X(u+z)}:\ , \qquad\quad
H_3(z)\ =\ \oint {du\over 4\pi} \del X(u+z)\ .
\efr
The allowed values of the chiral momenta are $p=n/2$,
and the simplest primary fields are the
$SU(2)$ highest weight states
$$
\psi_{J,J}(z)\ =\ e^{iJX}(z)
\efr
where $J=0, 1/2, 1, 3/2, \ldots$~.
By repeatedly acting with the lowering operator $H_-$, one builds up
a spin-$J$ $SU(2)$ multiplet of primary fields $\psi_{J,m}$, $m\in \{J,
J-1, \cdots, -J\}$, with conformal dimension $\Delta = J^2$
$$
\psi_{J,m}(z)\ = \left [ {(J+m)!\over (J-m)! (2J)!}\right ]^{1/2}
\left[\oint {du\over 2\pi i} : e^{-iX(z+u)}:
\right]^{J-m} \, \psi_{J,J}(z)\ .
\efr
After performing the contractions and the
contour integrals, $\psi_{J,m}$ become polynomials in the
derivatives of $X$.

Upon coupling to gravity, we dress the $c=1$ primaries
to obtain the physical operators with dimension one
$$
\Psi_{J,m}^\pm (z) \ =
\left [ (J+m)! (J-m)! (2J)!\right ]^{1/2}
\psi_{J,m}(z)\, :
e^{\epsilon^\pm_J \phi}(z) :
\nfr{dress}
where $\epsilon^\pm_J\ =\ -1 \pm J$, and the peculiar normalization
is chosen to simplify the subsequent calculation of the operator
algebra. It is easy to check that all the operators \dress\ satisfy
the Virasoro conditions of the $c=26$ $(X, \phi)$ CFT. The fact that
they are not pure gauge is obvious, because the corresponding states
have non-vanishing norms.

An important property of the chiral vertex operators
$\Psi^\pm_{J,m}$ is that they form an interesting
algebra under the O.P.E.$^{\Ref{W},\Ref{KP}}$
$$
\Psi^+_{J_1,m_1}(z)\, \Psi^+_{J_2,m_2}(0)\ =\ldots+\, {2\over z}
(J_1m_2 - J_2 m_1)\, \Psi^+_{J_1+J_2-1, m_1+m_2}(0)\, +\, \dots \ ,
\nfr{psiope}
where we have shown the only physical operator appearing on the
right-hand side.
This is isomorphic to a wedge sub-algebra of $w_\infty$.
Some of the consequences of this algebra will be explored in section
\S 4.

To construct the physical vertex operators of closed-string theory,
it is necessary to combine the $z$ and $\bar z$ sectors.
Thus, at the self-dual radius we obtain a class of operators of
the form
$ \Psi^\alpha_{J,m}(z) \cdot
\overline\Psi^\alpha_{J,m'}(\overline{z})$,
where $\alpha$ (not summed over) can be $+$ or $-$.
In fact, up to gauge equivalence, these are all the physical
operators without ghost excitations.~$^{\Ref{lz},\Ref{Bouw}}$
At any radius of compactification
we may construct combinations of the tachyon
operators and of the ``discrete primaries" appearing at the
self-dual radius with their anti-holomorphic
counterparts, that are permitted by the standard momentum restrictions.
For example, in the uncompactified
theory, in addition to the continuous spectrum of tachyons, we find
the discrete states
$ \Psi^\alpha_{J,m}(z) \cdot
\overline\Psi^\alpha_{J,m}(\overline{z})$, with $|m|<J$.
\section{BRST formalism and the ``Ground Ring"}
\noindent The structure of the discrete states' sector in the
two-dimensional string theory is richer than what we have seen up to
now because there are additional states with ghost excitations and
non-trivial ghost numbers. Such states can be studied only in the
framework of
the BRST quantization. There an important r\^ole is played by
the nihilpotent BRST charge~\note{For now, we are
restricting ourselves to the chiral sector.}
$$
Q_{BRST}\ =\ \frac1{2\pi i} \oint dz \, c(z)\left( T^{(X,\phi)}(z) +
\frac12 T^{(b,c)}(z) \right)\ .
\efr
Physical states are
given by the cohomology classes of the BRST
charge at ghost number one. In other words,
the physical operators are the ghost number one vertex operators
which commute with $Q_{BRST}$, modulo
commutators of $Q_{BRST}$ with other operators. In the BRST formalism,
the tachyon vertex
operators are given by
$$
V^\pm_k (z) \ =\ c(z)\, e^{ikX + (-1+\pm k)\phi}(z)
\efr
and the discrete vertex operators are
$$
Y^\pm_{J,m}(z)\ =\ c(z)\, \Psi^\pm_{J+1,m}(z)\ .
\nfr{ydef}
Since these operators are non-trivial only at the discrete values of
the momenta, it can be shown~$^{\Ref{W}}$
that each $Y$ has a partner cohomology class of adjacent ghost number.

Indeed, in the complete analysis of the BRST
cohomology~$^{\Ref{lz},\Ref{Bouw},\Ref{coho}}$ it was
found that the $Y^+_{J,m}$ have ``partners" at ghost number zero,
whereas the $Y^-_{J,m}$ have ``partners" at ghost number two. The ghost
number zero operators, usually denoted by ${\cal O}_{J,m}$,
form the ``{\sl Ground Ring\/}"~$^{\Ref{W}}$
$$ {\cal O}_{J,m}{\cal O}_{J',m'}={\cal O}_{J+J',m+m'}
\nfr{Gring}
which is to be interpreted as a fusion rule in BRST
cohomology (i.e. up to BRST commutators).
Their explicit form is given by
$$\eqalignno{&{\cal O}_{0,0}\ =\ 1 \cr
&{\cal O}_{\sssty {1\over 2},{1\over 2}}\ =\ (cb +\frac{i}2 \del X
-\frac12 \del\phi )\, e^{\frac12 (iX +\phi)} \cr
&{\cal O}_{\sssty{1\over 2},-{1\over 2}}\ =\ (cb -\frac{i}2 \del X
-\frac12\del\phi )\, e^{\frac12 (-iX +\phi)}
&\nameali{ringgen}\cr}
$$
The latter two are the ring generators which determine the form of the
entire ring via the fusion rules. One can think of the ring as
a set of analogues of the identity operator, which occur at the
discrete momenta. The presence of this ring, and other discrete
states, endows the two-dimensional string theory with many special
properties.

Witten~$^{\Ref{W}}$ has found a geometrical interpretation
of the $w_\infty$ currents and of the ground ring. The first step is to
note that the chiral ground ring can be interpreted as
polynomial functions on a two dimensional plane with coordinates $x =
{\cal O}_{\sssty{1\over 2},{1\over 2}}$,
$y = {\cal O}_{\sssty{1\over 2},-{1\over 2}}$,
i.e.  ${\cal O}_{J,m} \sim x^{J+m} y^{J-m}$. Next Witten studied
the action of the chiral currents
$\Psi^\pm_{J,m}$ on the chiral ground ring.
He found~$^{\Ref{W}}$ that the $\Psi^+_{J,m}$ generate the
area-preserving diffeomorphisms of the plane $x-y$ (a wedge
sub-algebra~$^{\Ref{wedge}}$ of
$w_\infty$).~\note{The algebra $w_\infty$ first
appeared in the matrix model formulation of 2-d quantum
gravity.~$^{\Ref{aj}}$} As an example, the first few
$\Psi^+_{J,m}$ operators act on the ground ring as follows,
$$\eqalignno{ &\Psi^+_{\sssty{1\over 2},{1\over 2}}\ \sim\
{\del\ \over \del y}
\qquad\qquad\qquad \Psi^+_{\sssty{1\over 2},-{1\over 2}}\ \sim\
{\del\ \over\del x}
\cr   &\Psi^+_{1,1}\ \sim\ x{\del\ \over\del y} \qquad\qquad\qquad
\Psi^+_{1,-1}\ \sim\ y{\del\ \over\del x}\cr
&\Psi^+_{1,0}\ \sim\ \frac12 \left(x{\del\ \over\del x} -
y{\del\ \over\del y}\right)\ . &\numali\cr}
$$
\jump

Now we consider the complete set of cohomology
classes of the chiral theory at the
self-dual radius. The operators ${\cal O}_{J,m}$ and $Y^+_{J,m}$
actually make up only half of the BRST
cohomology (also known as the relative
cohomology).~$^{\Ref{lz},\Ref{Bouw},\Ref{coho},\Ref{WZ}}$
This follows from the existence of the operator
$$
a\ =\ [Q_{BRST},\phi]\ = \ c\del\phi + 2\del c\ .
\efr
The operator $\phi$ is not a conformal field, but
$a$ is. This implies that in the usual space of conformal fields, $a$
is not BRST trivial (but it is BRST invariant). Thus, by applying $a$
to an operator, one gets a new BRST invariant vertex operator.
Following Ref. \Ref{WZ}, one defines
$$
a {\cal O}_{J,m}(0)\ =\ \frac1{2\pi i} \oint dz\, a(z) \cdot {\cal
O}_{J,m}(0)
\efr
and similarly for $aY^+_{J,m}$. $a{\cal O}_{J,m}$ is a new operator of
ghost number one, whereas $aY^+_{J,m}$ has ghost number two.
While ${\cal O}_{J,m}$ and
$Y^+_{J,m}$ are annihilated by $b_0 = \frac1{2\pi i} \oint dz \, z \,
b(z)$ (belong to relative cohomology),
$a{\cal O}_{J,m}$ and $aY^+_{J,m}$ are not. Thus we find the
following cohomology classes labeled by the ghost number $G$,
$$\eqalignno{
& G = 0 :\quad {\cal O}_{J,m} \cr
& G = 1 :\quad Y^+_{J,m} , \ a {\cal O}_{J,m}\cr
& G = 2 :\quad a Y^+_{J,m}\ . &\numali\cr}
$$
These states
are known to exhaust the absolute BRST cohomology for Liouville
energy $\epsilon>-1$. There are also their conjugate states
with $\epsilon<-1$ and ghost number $3-G$.
Denoting the operator conjugate to $Y^+_{J,m}$
by ${\cal P}_{J,m}$,  we find the complete list of cohomology
classes with Liouville energy $\epsilon<-1$,
$$\eqalignno{
& G = 1 :\quad Y^-_{J,m} \cr
& G = 2 :\quad {\cal P}_{J,m} , \ a Y^-_{J,m}\cr
& G = 3 :\quad a {\cal P}_{J,m}\ . &\numali\cr}
$$
\jump

Consider now the two dimensional closed string
theory. The closed string states
are annihilated by the full BRST charge and satisfy the
conditions $(L_0 - \overline{L}_0)\ket{\psi}=(b_0 -
\bar{b}_0)\ket{\psi}=0$.~\note{See Ref. \Ref{WZ}\ for a detailed
discussion of the $b_0 - \bar{b}_0$ condition.}
Witten and Zwiebach~$^{\Ref{WZ}}$ found that there
exist states which satisfy the $b_0 - \bar{b}_0$ condition but that
are not annihilated separately by $b_0$ and $\bar{b}_0$.
These states, which belong to the so-called semi-relative cohomology,
cannot be written as products of the holomorphic and anti-holomorphic
states. The complete list of $\epsilon>-1$
semi-relative cohomologies is~$^{\Ref{WZ}}$
$$\eqalignno{& G = 0 :\quad {\cal O}_{J,m}\overline{\cal O}_{J,m'}\cr
& G = 1 :\quad Y^+_{J,m} \overline{\cal O}_{J,m'}, \ {\cal O}_{J,m}
\overline{Y}^+_{J,m'},\ (a +\overline{a})\cdot ({\cal
O}_{J,m}\overline{\cal O}_{J,m'})\cr
& G = 2 :\quad Y^+_{J,m} \overline{Y}^+_{J,m'}, \ (a +\overline{a})\cdot
(Y^+_{J,m} \overline{\cal O}_{J,m'}), \ (a +\overline{a})\cdot
({\cal O}_{J,m} \overline{Y}^+_{J,m'})\cr
& G = 3 :\quad (a +\overline{a})\cdot (Y^+_{J,m}
\overline{Y}^+_{J,m'})\ .
&\nameali{clspone} \cr}
$$
The corresponding dual states with $\epsilon<-1$ are~$^{\Ref{WZ}}$
$$\eqalignno{& G = 2 :\quad Y^-_{J,m} \overline{Y}^-_{J,m'}\cr
& G = 3 :\quad Y^-_{J,m} \overline{\cal P}_{J,m'}, \ {\cal P}_{J,m}
\overline{Y}^-_{J,m'},\ (a +\overline{a})\cdot (Y^-_{J,m}
\overline{Y}^-_{J,m'})\cr
& G = 4 :\quad {\cal P}_{J,m}\overline{\cal P}_{J,m'},\
(a +\overline{a})\cdot (Y^-_{J,m} \overline{\cal P}_{J,m'}),
\ (a +\overline{a})\cdot({\cal P}_{J,m} \overline{Y}^-_{J,m'})\cr
& G = 5 :\quad (a +\overline{a})\cdot ({\cal P}_{J,m}
\overline{\cal P}_{J,m'}) \ . &\nameali{clsptwo} \cr}
$$
The dual states have ghost number $5-G$ because of the necessary
factor $b_0^-$ which we will discuss below. They can be interpreted as
Batalin-Vilkovisky anti-field vertex operators.~$^{\Ref{ver}}$

The $G=1$ vertex operators are of utmost importance because they give
rise to the currents that generate the infinite symmetry.
In the next section we will use these currents
to calculate certain correlation functions through the Ward identities.
We will make use only of those $G=1$ operators that can be factorized
into a product of chiral parts. The other operators,
$(a +\overline{a})\cdot ({\cal
O}_{J,m}\overline{\cal O}_{J,m'})$, which are not annihilated by
$b_0$, do not appear to produce interesting symmetries.
\note{Those operators that cannot be factorized into a
product of chiral parts play a rather special r\^ole in the theory.
Although they appear in the Ward identities, they seem to
be unnecessary in the on-shell formulation of the theory because
they correspond to auxiliary fields.~$^{\Ref{sakait},\Ref{mukher}}$
Nevertheless, they are clearly important in the B-V formulation.}
\chapter{Discrete states and Ward identities in 2-d closed
string theory}
\noindent In this section we will consider the uncompactified 2-d
closed string theory, making extensive use of the symmetries
generated by the $G=1$ operators
$$
Y^+_{J,m}(z) \overline{\cal O}_{J,m}(\bar{z})\ =\ c(z)\, W_{J,m}
(z,\bar{z})\ ,
\efr
where
$$
W_{J,m}(z,\bar{z})\ =\ \Psi^+_{J+1,m}(z)
\overline{\cal O}_{J,m}(\bar{z}) \ .
\efr
We will write the bulk correlation functions of
tachyons on a sphere,~\note{Correlation functions involving
the discrete vertex operators were considered in Refs. \Ref{WZ},
\Ref{Mats}, \Ref{Wu}, \Ref{Aref}, \Ref{Suz}, \Ref{Vdot}, \Ref{Ghos},
\Ref{pert}.}
$$
\vev{V^\pm_{k_1} \cdots V^\pm_{k_n} }\ ,
\efr
in two different ways. \newline
In the path integral formalism,
$$
\vev{V^\pm_{k_1} \cdots V^\pm_{k_n} }\ =\ \vev{ V^\pm_{k_1} (+\infty)
V^\pm_{k_2}(1) \int T^\pm_{k_3} \dots \int T^\pm_{k_{n-1}}\,
V^\pm_{k_n}(0) }
\nfr{anove}
where $\int T^\pm_k\ \buildchar{=}{\rm def}{ }\
\int d^2 y \, T^\pm_k(y,\bar{y})$. \newline
In the operator formalism,\note{See for example Refs.
\Ref{GSW}, \Ref{divec}\ and references therein.}
$$\eqalignno{
\vev{V^\pm_{k_1} \dots V^\pm_{k_n} }\ =\ &\bra{V^\pm_{k_1} }
\, V^\pm_{k_2}(1)\, \Delta\,  V^\pm_{k_3}(1)\, \Delta \cdots
\Delta\, V^\pm_{k_{n-1}}(1)\, \ket{V^\pm_{k_n}} \cr
& +\ {\rm permutations}&\nameali{adieci}\cr}
$$
where $\ket{V^\pm_{k_n}}\ \buildchar{=}{\rm def}{ }\
\lim_{z\rightarrow 0} V^\pm_{k_n}(z) \,\ket{0}$, $\bra{V^\pm_{k_1} }\
\buildchar{=}{\rm def}{ }\  \lim_{z\rightarrow\infty} \bra{0}\,
V^\pm_{k_1}(z)$. The propagator, including the
ghost contribution, is
$$\Delta\ =\ {b_0^+ b_0^- \over L_0+\bar L_0}\,\Pi_{L_0, \bar L_0}
\efr
where $b_0^\pm \ \buildchar{=}{\rm def}{ }\ b_0 \pm \bar{b}_0,$
and the Virasoro generators include the ghost pieces.
$\Pi_{L_0, \bar L_0}$ is the projector onto states that satisfy
$L_0=\bar L_0$.

In the following we will focus on the tachyon correlators of
type $(N, 1)$ and $(1, N)$, which are the only ones that do not vanish
for generic momenta. As we have seen in \S 2.2, these correlators are
given by the multiple integrals
$$ \eqalignno{A_{N, 1}(k_1, \ldots, k_N)\ &=\
\vev{ V^-_{-k} (\infty)
V^+_{k_2}(1) \int T^+_{k_3} \dots \int T^+_{k_{N-1}}\,
V^+_{k_N}(0) }\cr
&=\  \prod_l \int d^2 z_l \prod_i |z_i-1|^{4s_{i1}}
|z_i|^{4s_{i0}} \prod_{i<j} |z_i-z_j|^{4s_{ij}}\cr
&=\ {\pi^{N-2}\over (N-2)!} \prod_{r=1}^N
\frac{\Gamma(1-2k_r)}{\Gamma(2k_r)}
&\nameali{gen}\cr }
$$
where $l$, $i$ and $j$ run from 2 to $N-1$, and
$ s_{ij}=k_i k_j-\epsilon^+_i\epsilon^+_j=-1+k_i+k_j$.
The integrals of Eq. \gen\ are generalizations of the integrals
calculated in Ref. \Ref{Dots}\  by contour deformation techniques.
We will find that the integrals \gen\ can be calculated via simple
recursion relations which follow from the $w_\infty$ Ward identities.
This is a mathematical result which does not seem to be well known.
To accomplish this, we first need to review the systematic construction
of conserved currents and charges from the
discrete states.~$^{\Ref{WZ}}$
\section{BRST conserved currents and charges}
\noindent In two dimensions a conserved current is a pair $(J_z,
J_{\bar z})$ such that the one form
$$
\Omega^{(1)}\ =\ J_z dz - J_{\bar{z}} d \bar{z}
\efr
is closed, $ d \Omega^{(1)}=0=\overline\del J_z+\del J_{\bar z}$.
The associated conserved charge is ${\cal A}=\oint \Omega^{(1)}$.
Actually, we may demand that a less stringent condition is fulfilled.
In BRST quantization it is sufficient to require that
$$
d\Omega^{(1)}\ =\ \{Q_{BRST}, \Omega^{(2)}\} \qquad\quad {\rm or}
\qquad\quad \overline\del J_z+\del J_{\bar z}\ =\ -\{Q_{BRST},
\Omega^{(2)}_{z\bar{z}} \}
\efr
where $\Omega^{(2)}=\Omega^{(2)}_{z\bar{z}} \, dz \wedge d\bar{z}$.
Then ${\cal A}=\oint \Omega^{(1)}$ is conserved up to BRST trivial
(pure gauge) operators.

The condition that ${\cal A}$ be BRST invariant, $\{Q_{BRST}, {\cal
A}\}=0$, implies that there exists a zero form $\Omega^{(0)}$ such
that $d\Omega^{(0)} = \{Q_{BRST}, \Omega^{(1)}\}$. Moreover,
$\{Q_{BRST}, \Omega^{(0)}\}=0$ holds.

Thus, the existence of a conserved charge implies the descent
equations
$$\eqalignno{ 0 \ &=\ \{Q_{BRST}, \Omega^{(0)}\}\cr
d\Omega^{(0)}\ &=\ \{Q_{BRST}, \Omega^{(1)}\}\cr
d\Omega^{(1)}\ &=\ \{Q_{BRST}, \Omega^{(2)}\} \ .&\numali\cr}
$$
If we have found a $G=1$ BRST invariant
operator $\Omega^{(0)}$, these equations allow us to calculate the
corresponding $G=0$ current $\Omega^{(1)}$.
Taking
$$
\Omega^{(0)}_{J,m}(z,\bar{z})\ =\ c(z) W_{J,m}(z,\bar{z})
\ ,\nfr{adue}
the associated charges are found to be~$^{\Ref{WZ}}$
$$
{\cal A}_{J,m}\ =\ \oint {dz\over 2\pi i}\, W_{J,m}(z,\bar{z})
\ -\ \oint {d\bar{z}\over 2\pi i}\,
c(z) \Psi_{J+1,m}(z)
\overline{X}_{J,m}(\bar{z}),
\efr
where $\overline{X}_{J,m}(\bar{z}) =
\bar{b}_{-1}\overline{\cal O}_{J,m}(\bar{z})$.
{}From Eqs. \psiope\ and \Gring, the algebra of the charges is
$$
[{\cal A}_{J_1,m_1}{\cal A}_{J_2,m_2}]\ =2\bigl(
(J_1+1)m_2 - (J_2+1) m_1\bigr) {\cal A}_{J_1+J_2, m_1+m_2}\ .
\efr
Thus, the 2-d closed string has infinite symmetry isomorphic
to the area-preserving diffeomorphism. Note that in
the symmetry currents the usual chiral discrete states are sandwiched
with the ground ring operators in such a way that the left and right
momenta balance, producing a local quantum field. Thus, the presence
of both types of operators is crucial to the demonstration of the
infinite symmetry.

We may now consider the action of the charge ${\cal A}_{J,m}$ on $n$
tachyons.~$^{\Ref{K},\Ref{WZ}}$  Just from kinematical considerations,
we conclude~$^{\Ref{K}}$ that the charge
${\cal A}_{m+n-1,m}$ annihilates $l$ $T^+$ tachyons of generic
momenta, for $l<n$.
The first non-trivial action of this charge is on $n$
$T^+$ tachyons, producing only one $T^+$ tachyon:~$^{\Ref{K}}$
$$
{\cal A}_{m+n-1,m} \, V^+_{k_1}(0) \int T^+_{k_2} \dots \int T^+_{k_n}\
=\ F_{n,m}(k_1, \cdots ,k_n)\, V^+_k (0)
\nfr{action}
where $k= \sum k_i +m$.
A similar formula holds for ${\cal A}_{-m+n-1,m}$ acting on
$T^-$ tachyons.

The situation is more complex if one or more tachyons carry one of
the discrete momenta. For example,  let us consider the action of the
charges ${\cal A}_{m+n-1,m}$ on vertex operators $V^+_{s/2}$, with $s$
positive integers. For $n=1$ the action is correctly determined by
Eq. \action. However, for $n>1$, the charge does not annihilate the
vertex operator. Instead, it produces a discrete state in the
``semi-relative" cohomology.~$^{\Ref{WZ}}$
We will not analyze in detail such action of the charges in this paper.

In the following we will need the explicit expression of
$F_{n,m}(k_1, \cdots ,k_n)$. To compute it, consider first the action
of the charge ${\cal A}_{{\sssty 1\over
\sssty 2},-{\sssty 1\over\sssty 2}}$ on
two $T^+$ tachyons of generic momenta:
$$\eqalignno{
{\cal A}_{{\sssty 1\over\sssty 2},-{\sssty 1\over\sssty 2}}\,
V^+_{k_1} \int T^+_{k_2} =
&\left[\oint_\gamma {dz \over 2\pi i}\, W_{{\sssty 1\over\sssty 2},
-{\sssty 1\over\sssty 2}}
(z,\bar{z}) - \oint_\gamma {d\bar{z}\over 2\pi i} \, c(z)
\Psi_{{\sssty 3\over\sssty 2},-{\sssty 1\over\sssty 2}} (z)
\overline{X}_{{\sssty 1\over\sssty 2},-{\sssty 1\over\sssty 2}}
(\bar{z})\right] \cdot \cr
&\ \cdot\ V^+_{k_1}(0,0)\int d^2 w\, T^+_{k_2}(w,\bar{w})&\numali\cr}
$$
where the contour $\gamma$ encloses $0$. To get a
non-zero result from the action of the holomorphic part of the charge,
it is necessary that \noblackbox
$$
W_{{\sssty 1\over\sssty 2},-{\sssty 1\over\sssty 2}} (z,\bar{z})
V^+_{k_1}(0,0) \int d^2 w T^+_{k_2} (w,\bar{w}) = {1\over z}
F_{2,-\half}(k_1,k_2) V^+_{k_1+k_2
-{\sssty 1\over\sssty 2}}(0, 0)\ +\ \dots
\efr
The only contributions to the residue come from the
region where $z$ and $w$ approach each other and $0$. The
integral for $F_{2, -\half}(k_1,k_2)$ was calculated in Ref.
\Ref{K}\ using the results of Ref. \Ref{Kawai}\
$$
F_{2,-\half}(k_1,k_2)\ =\ 2\pi(2k_1+2k_2 -1)
{\Gamma(1-2k_1) \over \Gamma(2k_1)}
{\Gamma(1-2k_2) \over \Gamma (2k_2) } {\Gamma(2k_1+2k_2-1) \over
\Gamma (2-2k_1-2k_2) }\ .
\efr
Also, in Ref. \Ref{WZ}\ it was shown that the
anti-holomorphic part of the charge gives a contribution proportional
to $\bar{z}^s$ with $s> -1$. This is not singular enough to be relevant
and, therefore, the anti-holomorphic part of the charge does not
contribute in this case. One could anticipate this on general grounds
because this part of the charge carries (left,right) ghost numbers
equal to $(1, -1)$. The conservation of these ghost numbers forbids
the physical tachyon from appearing in the {\scaps o.p.e.}~.

In an analogous way one can compute the action of the charge ${\cal
A}_{m,m}$ on one $T^+_k$ tachyon (see Ref. \Ref{K}). Using these
results and the $w_\infty$ algebra of charges, in Ref. \Ref{K}\
the action of the charge ${\cal A}_{m+n-1,m}$ on $n$ $T^+$ tachyons
was determined. The final result for Eq. \action\ is \yesblackbox
$$
F_{n, m}(k_1, \dots, k_n)\ = \ 2\pi^{n-1} (n!)\, k
{\scriptstyle\Gamma(2k) \over \scriptstyle\Gamma(1-2k)}\,
\prod_{i=1}^n
{\scriptstyle\Gamma(1-2k_i) \over \scriptstyle\Gamma(2k_i)}\,
\nfr{explicit}
where $k=m+\sum_{i=1}^n k_i$ and $n\geq 1$.
A similar formula obviously holds for ${\cal A}_{-m+n-1,m}$
applied to $n$ generic $T^-$ vertex operators.
Note that an explicit evaluation
of this formula would require performing $n-1$ integrals, a very
difficult task, which is avoided here thanks
to the algebraic structure of the model.
\section{$w_\infty$ Ward identities}
\noindent
Perhaps, the simplest statement of the Ward identities~\note{For
another approach to the Ward identities see Refs. \Ref{KMS},
\Ref{Kachru}.} is through the
observation~$^{\Ref{W},\Ref{Polyakov}}$ that the theory possesses
states that are both pure gauge and identically zero. The vertex
operator for such a state is
$$\{Q_{BRST}, cW_{J, m}\}=0\nfr{uno}
and formally carries the physical ghost number $n_{gh}=2$.
It vanishes identically because the zero-picture current $cW_{J, m}$
is BRST invariant. A special feature of the theory under consideration
is the presence of an infinite number of such BRST invariant
zero-picture currents of ghost number 1. In general, linearized closed
string gauge invariance assumes the form
$$\delta \Psi=\{Q_{BRST}, \lambda\}
\nfr{gauge}
where the ghost numbers of $\Psi$ and $\lambda$ are 2 and 1
respectively. Therefore, there is an infinite number of gauge
parameters, $\lambda=cW_{J, m}$, for which the linearized gauge
invariance is trivial. In fact, as stated in Refs.
\Ref{W}, \Ref{Polyakov}, the presence of such trivial gauge
transformations guarantees that there are discrete states of ghost
number 2 which cannot be gauged away. While at a general momentum
there are enough gauge invariances to gauge away all the oscillator
states, at the discrete momenta carried by $cW_{J, m}$ some gauge
invariances become trivial, Eq. \uno, and there appear physical
oscillator states that cannot be gauged away. Thus, in a theory
with generally continuous momenta, the presence of symmetry charges
seems intrinsically connected with the presence of discrete states,
which are physical only at special discrete momenta.
In higher-dimensional theories the only known cases of
this phenomenon occur at zero momentum.
Some familar examples are the conservation of charge and the
associated extra photon state at zero momentum, and the conservation
of momentum and the extra graviton-dilaton states at zero momentum.

The Ward identities follow after inserting Eq. \uno\  into
correlation functions.~$^{\Ref{WZ},\Ref{ver}}$ As usual, the BRST
anti-commutator can be re-written as a sum over the boundaries of
the moduli space of a sphere with $n$ punctures where the sphere
is pinched into two spheres with $m$ and $n-m$ punctures respectively.
In the literature on string theory these boundaries of moduli space
are sometimes referred to as the ``canceled propagators''. The reason
for this terminology is most apparent in the operator formalism.
In the 26-dimensional string theory, the
``canceled propagator argument'' is the statement that such boundary
terms on moduli space typically vanish (at least in the context of
an appropriate analytic continuation), because the momentum that
flows through the pinch is off-shell. In the 2-dimensional string
theory the situation is very different. As emphasized in Ref.
\Ref{K}, there is an infinite set of special kinematical
arrangements where the momentum that flows through the pinch is
precisely on-shell. When this is the case, the ``canceled propagator
contribution'' is non-vanishing and explicitly calculable.
The Ward identity is nothing but the statement that the sum of all the
canceled propagator contributions, which arise after the insertion of
Eq. \uno\ into a correlation function, vanishes. We emphasize
that one can search for the non-trivial contributions to the Ward
identity simply on the basis of studying the kinematics:
whenever the momentum that flows through the canceled propagator
corresponds to an on-shell state, one expects, and usually finds, a
non-vanishing contribution.

Thus the Ward identities for tachyon correlation functions assume the
form
$$
\vev{\{Q_{BRST} , cW_{J, m}\}V^\pm_{k_1} \cdots V^\pm_{k_n} }\ =\ 0\ .
\efr
More explicitly, in the operator formalism we have
$$\eqalignno{
0\ =\ &\bra{V^\pm_{k_1} } \{Q_{BRST} , cW_{J, m}\} (1)\, \Delta \,
V^\pm_{k_2}(1)\,\Delta \cdots  \Delta\, V^\pm_{k_{n-1}}(1)\,
\ket{V^\pm_{k_n}} \cr
&\ + \ {\rm permutations}\ .&\nameali{auno}\cr}
$$
Having written the Ward identity in this form, we can now apply the
``canceled propagator argument". In the operator formalism this is
simply obtained by explicitly commuting $Q_{BRST}$ through
each of the propagators $\Delta$. First observe that
$$
\left[ Q_{BRST} , \Delta \right] \ =\ \Pi_{L_0, \bar L_0}\ b_0^- \
=\ \sum_i\ket{\Phi_i}\bra{\Phi^i}\, b_0^-
\nfr{bzero}
where the sum is over a complete set of states that satisfy
$(L_0-\bar L_0)\ket{\Phi_i}=0$.
Thus, $Q_{BRST}$ can literally cancel a propagator, replacing it with
an insertion of $b_0^-$. The conjugate states
$\ket{\Phi_i}$ and $\ket{\Phi^i}$ satisfy $\bivev{\Phi^j}{\Phi_i}
= \delta^j_i$ (for continuous spectrum the Kroenecker symbol is
replaced by the Dirac delta function).
Each pair of conjugate states has their ghost numbers
add up to 6, their momenta $k$ add up to 0, and their energies
$\epsilon$ add up to $-2$. For instance, the state conjugate to
$\ket{V^\pm_k}$ is
$$\ket{\v^\pm_{-k}}\ =\ c \partial c\,\bar c \bar \partial\bar c
\, T^\pm_{-k}(0) \ket{0}\ .
\efr
It is also convenient to define the states $\ket{\widetilde{V}^\pm_k}
\ \buildchar{=}{\rm def}{ }\ b_0^-\ \ket{\v^\pm_k}$
which have ghost number 3 and are annihilated by
$b_0^-$.~$^{\Ref{WZ},\Ref{ver}}$
In the Batalin-Vilkovisky quantization~$^{\Ref{bv}}$ these are
to be thought of as ``anti-tachyons''.~$^{\Ref{ver}}$

Now, commuting $Q_{BRST}$ in Eq. \auno\ with the vertices and
propagators until it annihilates against the vacua, one gets a sum of
``canceled propagator contributions''. Each one has the form
$$
\sum_i\bra{V_{k_1}} V_{k_2} \Delta \cdots \Delta
V_{k_p}\ket{\Phi_i} \bra{\widetilde \Phi^i} V_{k_{p+1}}
\Delta cW_{J, m} \Delta \cdots V_{k_{n-1}} \ket{V_{k_n}} \ ,
\nfr{canp}
where $\ket{\widetilde\Phi^i}=b_0^-\ket{\Phi^i}$. By ghost number
counting, $\ket{\Phi_i}$ must have the physical ghost number 2, and,
therefore, $\ket{\widetilde\Phi^i}$ carries ghost number 3. Also,
the energy and momentum of $\ket{\Phi_i}$ are completely determined
because the correlation functions obey the sum rules Eq. \sumrules.
It may happen that at this energy and momentum there is no physical
state. Then the usual ``canceled propagator argument'' applies
and we conclude that Eq. \canp\ vanishes. In this theory
there is a class of cases, however, where there are physical states
contributing to the sum over $i$. Here we will only analyze the
situations where the energy and momentum that flow through the
``canceled propagator'' obey the tachyon dispersion relation.
Then the sum over $i$ collapses to one non-vanishing term involving
the tachyon states.

Note that Eq. \canp\ can be interpreted as the
pinching of a sphere with $n$ punctures into
two spheres with $p$ and $n-p$ punctures respectively.
The amplitude corresponding to one of these spheres is perfectly
conventional, with each vertex operator carrying ghost number 2.
The other amplitude is unusual, since the current
operator $cW_{J, m}$ carries ghost number 1, while
$\ket{\widetilde\Phi^i}$ carries ghost number 3. In the next
section we will
show that this amplitude can be interpreted as a matrix
element of the charge operator ${\cal A}_{J, m}$.
For the correlation functions involving a current
we introduce the notation
$$\vev{\widetilde V_{-k} V_{k_1}\cdots V_{k_n} cW_{J, m}}=
\sum_P \bra{\widetilde V_{-k}} V_{k_1} \Delta cW_{J, m}
\Delta \cdots V_{k_{n-1}} \ket{V_{k_n}} \ .
\nfr{unusual}
The permutations $P$ involve all the vertex operators, except for
$\widetilde V$ which has to be kept as an out-state (in the
leftmost position). Since $\widetilde V$ and $cW$ are both
anti-commuting, this definition guarantees that all the
permutations contribute with the same sign. In general, in
dealing with correlators involving anti-commuting
vertex operators, their ordering becomes important, and
one may have to appeal to the operator formalism in order to
construct their proper definition.

Taking into account the sum over all permutations in Eq. \auno,
the final form of the Ward identity is~$^{\Ref{ver}}$
$$0\ =\ \sum_{{\cal I}, {\cal L}}
\vev{V_{i_1} \dots V_{i_I} \Phi}
\vev{\widetilde\Phi\, V_{l_1}\cdots V_{l_L}\, cW_{J, m}}
\nfr{aund}
where the sum is over all possible partitions of the vertex operators
$V$ into two sets ${\cal I}$ and ${\cal L}$ with $I$ and $L$ elements
respectively.

Now, as we remarked earlier, there are cases when the kinematics
restricts $\Phi$ to be on the tachyon mass shell. Let us analyze
the energy and momentum conservation laws for the correlator
$$\vev{\widetilde\Phi\, V^+_{k_1}\cdots V^+_{k_n}\, cW_{J, m}}\ .
\nfr{ccorr}
If $\widetilde\Phi$ carries momentum $P$ and energy $E$, we obtain
$$\eqalign{&m+P+\sum_{i=1}^n k_i =0\cr
& J+E +\sum_{i=1}^n (-1+k_i) =-2\ .\cr}
\efr
For the number of particles
$n=J-m+1$, it follows that $E=-1+P$, which is the dispersion
relation of a positive chirality tachyon.
Remarkably, this applies for arbitrary momenta $k_i$, which is a
consequence of the two-dimensional kinematics!
Therefore, whenever $n=J-m+1$, Eq. \aund\
receives a contribution from the canceled propagator with
${\widetilde \Phi}={\widetilde V^+_P}$, $\Phi= V^+_{-P}$.
Similarly, if we flip all the
chiralities, then the same situation arises for $n=J+m+1$. It is not
hard to see that these two cases are the only non-trivial amplitudes
of type \ccorr\ involving tachyons of generic (not discrete) momenta.
\section{Correlation functions from the Ward identities}
\noindent
We will now use the Ward identities of Eq. \aund\
to calculate the correlation functions of
tachyons explicitly.
As an intermediate step, we can use the $w_\infty$ algebra
to calculate the ``unusual" correlators \unusual . In
Ref. \Ref{KPas}\ they were reduced to the action of charges on tachyons,
$$\eqalignno{&
\vev{\widetilde V^+_{P}(\infty) \, c(w) W_{J,m}(w,\bar{w}) V^+_{k_1}
(0) \int T^+_{k_2} \dots \int T^+_{k_n}}\ = \qquad\qquad
&\nameali{asei}\cr &\qquad\qquad\qquad\qquad\qquad =\
\bivev{\v^+_{P}}{\,{\cal A}_{J,m} V^+_{k_1} (0)
\int T^+_{k_2} \dots \int T^+_{k_n}} \cr}
$$
where the charge operator acts on all the tachyons to its right.
{}From Eq. \action\ it immediately follows that
$$\vev{\widetilde V^+_{-k} V^+_{k_1} V^+_{k_2} \cdots
V^+_{k_n} cW_{m+n-1, m}}\ =\  F_{n, m}(k_1, \dots , k_n)
\nfr{unusres}
where $k=m+\sum_{i=1}^n k_i$, and $F_{n,m}$ is given by Eq. \explicit.

Now we can calculate all the correlation functions
$A_{N,1}$, given by equation
\gen, from the Ward identity
$$\vev{\{Q_{BRST}, cW_{-m, m}\}
V^+_{k_1}\cdots V^+_{k_N} V^-_{-\half}}=0
\nfr{newward}
where $N\geq 3$, and the kinematics fixes $m=1-\frac{N}2$.
Summing over all the non-vanishing canceled propagators, we find
$$\eqalignno{ 0\ =\ &\sum_{i=1}^N \vev{\widetilde V^+_{-k}
V^+_{k_{p(1)}} V^+_{k_{p(2)}} \cdots V^+_{k_{p(N-1)}}\,
cW_{-m, m}} \vev{V_k^+\, V^+_{k_i}
V^-_{-{\sssty 1\over\sssty 2}}} \cr
&\ +\ \vev{\widetilde V_{-r}^- \, V^-_{-{\sssty 1\over\sssty 2}}
cW_{-m, m}} \vev{V_{r}^-\, V^+_{k_1}
V^+_{k_2} \cdots V^+_{k_{N-1}}\, V^+_{k_N} }
&\nameali{Ward} \cr}
$$
where $p$ is the set of the first $N$ strictly positive integers except
for $i$, and $p(j)$ is the $j^{\rm th}$ element of the set.
The momenta of the intermediate
states, $r= \frac{1-N}2$ and $k= 1-\frac{N}2 + \sum_{j=1}^{N-1}
k_{p(j)}$, have been fixed
using the momentum and energy conservation laws.

Now we use Eq. \unusres, where $F_{n, m}$ is given by Eq. \explicit, to
substitute the explicit expressions for the correlators involving the
current. Thus, we obtain a linear relation expressing the tachyon
$N+1$-point function in terms of the tachyon three-point function,
$$\eqalignno{
[(N-2)!]^2 (N-1) \, \cdot\, & A_{N,1}(k_1, \dots, k_N)
\ = &\nameali{all}\cr
&=\  \sum_{i=1}^N
F_{N-1, m} (k_{p(1)}, \dots , k_{p(N-1)} )\, \cdot\,
A_{2,1}(k, k_i )\ .\cr}
$$
Using the momentum conservation equation $2 \sum_{j=1}^N k_j =(N-1)$,
one gets
$$
F_{N-1, m} (k_{p(1)}, \dots , k_{p(N-1)} )\ =\
\pi^{N-2} (N-1)!\, (1-2k_i) \prod_{l=1}^N
\frac{\Gamma(1-2k_l)}{\Gamma(2k_l)} \ .
\efr
The three-point function $A_{2, 1}$ contains no integrations.
Therefore, it is independent of the momenta and is normalized as
$A_{2,1}=1$. Now Eq. \all\ gives
$$
[(N-2)!]^2 (N-1) \, \cdot\, A_{N,1}(k_1, \dots, k_N)\ =\
\pi^{N-2} (N-1)!\,  \prod_{l=1}^N
\frac{\Gamma(1-2k_l)}{\Gamma(2k_l)}
\efr
from which we recover the correct answer Eq. \gen
$$
A_{N,1}(k_1, \dots, k_N)\ =\
{\pi^{N-2}\over (N-2)!} \prod_{l=1}^N
\frac{\Gamma(1-2k_l)}{\Gamma(2k_l)}\ .
\efr
\chapter{Conclusions}
\noindent The emergence of the $w_\infty$ symmetry structure in
two-dimensional string theory is an interesting result, whose deep
reason remains somewhat obscure. It may prove useful in elucidating
the connection between the conventional path integral techniques
and the matrix models, where the free fermions miraculously appear
and also give rise to $w_\infty$ symmetry. Despite some recent progress,
this connection is yet far from complete. In particular, we are
missing a precise dictionary translating the matrix model objects into
those of the continuum approach. If such a dictionary is found, it
will undoubtedly provide us with new insight into closed string theory.
\acknowledgements
We thank C.~Nappi, A.~Schwimmer, E.~Verlinde,
H.~Verlinde, E.~Witten, B.~Zwiebach and S.~Yankielowicz for helpful
discussions. I. R. K. is especially indebted to A. Polyakov for
collaboration and many illuminating discussions.
The research of I. R. K. is supported in part by
DOE grant DE-AC02-76WRO3072, NSF Presidential Young Investigator
Award PHY-9157482, James S. McDonnell Foundation grant No. 91-48, and
an A.P. Sloan Foundation Research Fellowship. The research of A.P. is
supported by an I.N.F.N. fellowship and partially by the NSF grant
PHY90-21984.
\references
\beginref
\Rref{Kawai}{H.~Kawai, D.~Lewellen and S.H.~Tye,
{\sl Nucl. Phys.} {\bf B269} (1986) 1.}
\Rref{GMil}{D.J.~Gross and N.~Miljkovic, {\sl Phys. Lett.}
{\bf 238B} (1990) 217;
\newline E.~Brezin, V.A.~Kazakov and Al.B.~Zamolodchikov,
{\sl Nucl. Phys.} {\bf B338} (1990) 673;\newline
P.~Ginsparg and J.~Zinn-Justin,
{\sl Phys. Lett.} {\bf B240} (1990) 333;\newline
G.~Parisi, {\sl Phys. Lett.} {\bf B238} (1990) 209, 213.}
\Rref{GKleb}{D.J.~Gross and I.R.~Klebanov,
{\sl Nucl. Phys.}  {\bf B344} (1990) 375.}
\Rref{I}{I. R. Klebanov, ``{\sl String Theory in Two Dimensions\/}",
PUPT-1271, in {\sl String Theory and Quantum Gravity '91},
World Scientific 1992.}
\Rref{vk}{V. Kazakov, in {\sl Random Surfaces and Quantum Gravity},
O. Alvarez et al. eds.~. }
\Rref{discrete}{J.~Goldstone, unpublished;
V.G.~Kac, in {\sl Group Theoretical Methods in Physics\/},
Lecture Notes in Physics, vol. 94, Springer-Verlag, 1979.}
\Rref{dj}{S. Das and A. Jevicki, {\sl Mod. Phys. Lett.} {\bf A5} (1990)
1639.}
\Rref{GKN}{D.J.~Gross, I.R.~Klebanov and M.J.~Newman,
{\sl Nucl. Phys.} {\bf B350} (1991) 621;\newline
D.J. Gross and I.R. Klebanov, {\sl Nucl. Phys.} {\bf B352} (1991) 671;
\newline K. Demeterfi, A. Jevicki and J.P. Rodrigues, {\sl Nucl.
Phys.} {\bf B362} (1991) 173, {\bf B365} (1991) 499, {\sl Mod. Phys.
Lett.} {\bf A6} (1991) 3199; \newline
U.H. Danielsson and D.J. Gross, {\sl Nucl. Phys.} {\bf B366} (1991) 3.}
\Rref{lz}{B.~Lian and G.~Zuckerman, {\sl Phys. Lett.} {\bf 254B}
(1991) 417, {\bf 266B} (1991) 21.}
\Rref{aj}{J.~Avan and A.~Jevicki,
{\sl Phys. Lett.} {\bf 266B} (1991) 35, {\bf 272B} (1991) 17,
{\sl Mod. Phys. Lett.} {\bf A7} (1992) 357; preprints
BROWN-HET-847 and 869;\newline
D.~Minic, J.~Polchinski and Z.~Yang,
{\sl Nucl. Phys.} {\bf B369} (1992) 324;\newline
G.~Moore and N.~Seiberg, {\sl Int. Jour. Mod. Phys.} {\bf A7} (1992)
2601;  \newline
S.~Das, A.~Dhar, G.~Mandal and S.~Wadia,
preprints
IASSNS-HEP-91/52, 91/89; {\sl Mod. Phys. Lett.} {\bf A7} (1992)
71, 937, 2245.}
\Rref{Dots}{Vl.S.~Dotsenko and V.A.~Fateev,
{\sl Nucl. Phys.} {\bf B251} (1985) 291.}
\Rref{Sasha}{A.M.~Polyakov, {\sl Mod. Phys. Lett.} {\bf A6} (1991) 635.}
\Rref{GK}{D.J.~Gross and I.R.~Klebanov, {\sl Nucl. Phys.} {\bf B359}
(1991) 3.}
\Rref{kdf}{P.~DiFrancesco and D.~Kutasov,
{\sl Phys. Lett.} {\bf 261B} (1991) 385, {\sl Nucl. Phys.} {\bf B375}
(1992) 119.}
\Rref{K}{I.R.~Klebanov, {\sl Mod. Phys. Lett.} {\bf A7} (1992) 723.}
\Rref{W}{E.~Witten, {\sl Nucl. Phys.} {\bf B373} (1992) 187.}
\Rref{KP}{I.R.~Klebanov and A.M.~Polyakov,
{\sl Mod. Phys. Lett.} {\bf A6} (1991) 3273.}
\Rref{Polyakov}{A.M.~Polyakov, ``{\sl Singular states in 2d quantum
gravity\/}", preprint PUPT-1289, September 1991.}
\Rref{WZ}{E.~Witten and B.~Zwiebach, {\sl Nucl. Phys.} {\bf B377}
(1992) 55.}
\Rref{ver}{E.~Verlinde, {\sl Nucl. Phys.} {\bf B381} (1992) 141.}
\Rref{KMS}{D.~Kutasov, E.~Martinec and N.~Seiberg, {\sl Phys. Lett.}
{\bf 276B} (1992) 437.}
\Rref{Kachru}{S.~Kachru, {\sl Mod. Phys. Lett.} {\bf A7} (1992) 1419.}
\Rref{divec}{P.~di~Vecchia, R.~Nakayama, J.L.~Petersen and S.~Sciuto,
{\sl Nucl. Phys.} {\bf B282} (1987) 103;
\newline
P.~di~Vecchia, M.~Frau, A.~Lerda and S.~Sciuto,
{\sl Nucl. Phys.} {\bf B298} (1988) 526.}
\Rref{bv}{I.~Batalin and G.~Vilkovisky,
{\sl Phys. Lett.} {\bf 102B} (1981) 27, {\bf 120B} (1983) 166;\newline
see also M. Henneaux, ``{\sl Lectures on the antifield-BRST formalism
for gauge theories\/}", proceeding of the XX GIFT meeting, and
M.~Henneaux and C.~Teitelboim, {\sl Quantization of gauge systems\/},
to be published by Princeton University Press.}
\Rref{GSW}{M.~Green, J.~Schwarz and E.~Witten, {\sl Superstring
Theory\/}, Vol. I and II, Cambridge University Press, Cambridge,
1987.}
\Rref{Liouv}{A.M.~Polyakov, {\sl Phys. Lett.} {\bf 103B} (1981) 207.}
\Rref{ddk}{F.~David, {\sl Mod. Phys. Lett.} {\bf A3} (1988)
1651;\newline
J.~Distler and H.~Kawai, {\sl Nucl. Phys.} {\bf B321} (1989) 509.}
\Rref{mxmod}{V. Kazakov, {\sl Phys. Lett.} {\bf 150B} (1985) 282;
\newline F. David, {\sl Nucl. Phys.} {\bf B257} (1985) 45; \newline
J. Ambjorn, B. Durhuus and J. Frolich, {\sl Nucl. Phys.} {\bf B257}
(1985) 433;\newline
V. Kazakov, I. Kostov and A. Migdal, {\sl Phys. Lett.} {\bf 157B}
(1985) 295.}
\Rref{GM}{ D.J. Gross and A. Migdal, {\sl Phys. Rev. Lett.} {\bf 64}
(1990) 127, 717, {\sl Nucl. Phys.} {\bf B340} (1990) 333;\newline
M. Douglas and S. Shenker, {\sl Nucl. Phys.} {\bf B335} (1990)
635;\newline
E. Brezin and V. Kazakov, {\sl Phys. Lett.} {\bf 326B} (1990)
144.}
\Rref{Doug}{M. Douglas, {\sl Phys. Lett.} {\bf 238B} (1990) 176.}
\Rref{Gervais}{T.L. Curtright and C.B. Thorn, {\sl Phys. Rev. Lett.}
{\bf 48} (1982) 1309;\newline
E. Braaten, T.L. Curtright and C.B. Thorn, {\sl Phys. Lett.} {\bf
118B} (1982) 115, {\sl Ann. Phys.} {\bf 147} (1983) 365;\newline
J.-L. Gervais and A. Neveu, {\sl Nucl. Phys.} {\bf B199} (1982) 59,
{\bf B209} (1982) 125, {\bf B224} (1983) 329, {\bf B238} (1984) 125,
396, {\sl Phys. Lett.} {\bf 151B} (1985) 271.}
\Rref{dhoker}{N. Mavromatos and J. Miramontes, {\sl Mod. Phys. Lett.}
{\bf A4} (1989) 1849;\newline
E. D'Hoker and P.S. Kurzepa, {\sl Mod. Phys. Lett.} {\bf
A5} (1990) 1411;\newline
E. D'Hoker, {\sl Mod. Phys. Lett.} {\bf A6} (1991) 745.}
\Rref{Bouw}{P. Bouwknegt, J. McCarthy and K. Pilch, {\sl Comm. Math.
Phys.} {\bf 145} (1992) 541, ``{\sl Semi-infinite cohomology in
conformal field theory and 2d gravity\/}", preprint USC-92/020,
ADP-92-194/M12, CERN-TH.6646/92, hep-th/9209034, September 1992.}
\Rref{coho}{S. Mukherji, S. Mukhi and A. Sen, {\sl Phys. Lett.} {\bf
266} (1991) 337;\newline
K. Itoh and N. Ohta, {\sl Nucl. Phys.} {\bf B377} (1992) 113, ``{\sl
Spectrum of two-dimensional (Super) Gravity\/}", preprint OS-GE-22-91,
hep-th/9201034, September 1991;\newline
C. Imbimbo, S. Mahapatra and S. Mukhi, {\sl Nucl. Phys.} {\bf B375}
(1992) 399.}
\Rref{KPas}{I.R. Klebanov and A. Pasquinucci, ``{\sl Correlation
functions from two-dimensional string Ward identities\/}", preprint
PUPT-1313, hep-th/9204052, April 1992.}
\Rref{wedge}{I. Bakas, {\sl Phys. Lett.} {\bf 228B} (1989) 57;\newline
C. Pope, L. Romans and X. Shen, {\sl Nucl. Phys.} {\bf B339} (1990)
191;\newline E. Bergshoeff, M.P. Blencowe and K.S. Stelle, {\sl Comm.
Math. Phys.} {\bf 128} (1990) 213.}
\Rref{Mats}{Y. Matsumura, N.Sakai and Y. Tanii, {\sl Coupling of
tachyons and discrete states in $c=1$ 2D gravity\/}", preprint
TIT/HEP-186, STUPP-92-124, hep-th/9201065, ``{\sl
Interaction of tachyons and discrete states in $c=1$ 2D quantum
gravity\/}, preprint TIT/HEP-187, hep-th/9201066, January
1992.}
\Rref{Wu}{Y.-S. Wu and C.-J. Zhu, ``{\sl The complete structure of the
cohomology ring and associated symmetries in D=2 string theory\/}",
preprint UU-HEP-92/6, June 1992.}
\Rref{Ghos}{D. Ghoshal, D.P. Jatkar and S. Mukhi, ``{\sl Kleinan
singularities and the ground ring of $c=1$ string theory\/}', preprint
TIFR/TH/92-34, June 1992.}
\Rref{Suz}{N.Ohta and H. Suzuki, ``{\sl Interaction of discrete states
with non zero ghost number\/}", preprint OS-GE 25-92, hep-th/9205101,
May 1992.}
\Rref{sakait}{N. Sakai, Y. Tanii, ``{\sl Physical Degrees of
Freedom in 2-D String Theories\/}", preprint TIT/HEP-203,
STUPP-92-129, July 1992.}
\Rref{mukher}{S. Mahapatra, S. Mukherji and A. M. Sengupta,
``{\sl Target space interpretation of new moduli in 2D string
theory\/}", preprint TIFR/TH/92-30, hep-th/9206111, June 1992.}
\Rref{Vdot}{Vl.S. Dotsenko, {\sl Mod. Phys. Lett.} {\bf A7} (1992)
2505.}
\Rref{Aref}{I.Ya. Aref'eva and A.P. Zubarev, ``{\sl Divergences of
discrete states amplitudes and effective lagrangians in 2D string
theory\/}", preprint hep-th/9205020, May 1992.}
\Rref{Polch}{J. Polchinski, {\sl Nucl. Phys.} {\bf B346} (1990) 253.}
\Rref{pert}{M.~Li, {\sl Nucl. Phys.} {\bf B382} (1992) 242;
\newline
J.~Barbon, ``{\sl Perturbing the ground ring of 2-d string theory\/}",
preprint CERN-TH6379-92, January 1992.}
\Rref{ns}{N. Seiberg, {\sl Prog. Theor. Phys. Suppl.} {\bf 102} (1990)
319.}
\endref
\ciao
